\documentclass[aps,nofootinbib,preprintnumbers]{revtex4}
\usepackage{lipsum}
\usepackage{CJK}
\usepackage{amsfonts}
\usepackage{amsmath}
\usepackage{amssymb}
\usepackage[english]{babel}
\usepackage{graphicx}
\usepackage{epsfig}
\usepackage{bm}
\usepackage{verbatim}
\usepackage[utf8]{inputenc}
\usepackage{booktabs}
\usepackage{multirow}
\usepackage{subfig}
\usepackage{slashed}
\usepackage{xcolor}
\usepackage[colorlinks=true,urlcolor=red,citecolor=red]{hyperref}
\usepackage[font=small]{caption}
\usepackage{float}
\usepackage{blindtext}
\usepackage{placeins}
\usepackage[utf8]{inputenc}
\usepackage{bbm}
\usepackage[colorinlistoftodos]{todonotes}

\newenvironment{Eqnarray}{\arraycolsep 0.14em\begin{eqnarray}}{\end{eqnarray}}
\newcommand{\ba}{\begin{Eqnarray}}
\newcommand{\ea}{\end{Eqnarray}}
\newcommand{\be}{\begin{equation}}
\newcommand{\ee}{\end{equation}}
\newcommand{\bal}{\begin{aligned}}
\newcommand{\eal}{\end{aligned}}
\newcommand{\bea}{\begin{eqnarray}}
\newcommand{\eea}{\end{eqnarray}}
\newcommand{\ben}{\begin{enumerate}}
\newcommand{\een}{\end{enumerate}}
\newcommand{\bit}{\begin{itemize}}
\newcommand{\eit}{\end{itemize}}
\newcommand{\bde}{\begin{widetext}}
\newcommand{\ede}{\end{widetext}}

\renewcommand{\[}{\left[}

\def\lsim{\mathrel{\rlap{\lower4pt\hbox{\hskip1pt$\sim$}}
    \raise1pt\hbox{$<$}}}
\def\gsim{\mathrel{\rlap{\lower4pt\hbox{\hskip1pt$\sim$}}
    \raise1pt\hbox{$>$}}}

\def\3211{$\mathrm{SU(3) \otimes SU(2)_L \otimes U(1)_R \otimes U(1)_{B-L}}$ }
\def\321{$\mathrm{SU(3) \otimes SU(2) \otimes U(1)}$ }
\def\422{$\mathrm{SU(4) \otimes SU(2) \otimes SU(2)_R}$ }

\newcommand{\mathsym}[1]{{}}

\definecolor{bostonuniversityred}{rgb}{0.8, 0.0, 0.0}

\input{tcilatex}
\begin{document}

\title{Modular $S_{3}$ flavoured Pati-Salam model with two family seesaw}
\author{A. E. C\'arcamo Hern\'andez$^{a,b,c}$}
\email{antonio.carcamo@usm.cl}
\affiliation{$^{{a}}$Universidad T\'ecnica Federico Santa Mar\'{\i}a, Casilla 110-V,
Valpara\'{\i}so, Chile\\
$^{{b}}$Centro Cient\'{\i}fico-Tecnol\'ogico de Valpara\'{\i}so, Casilla
110-V, Valpara\'{\i}so, Chile\\
$^{{c}}$Millennium Institute for Subatomic Physics at the High-Energy
Frontier, SAPHIR, Calle Fern\'andez Concha No 700, Santiago, Chile}

\author{Ivo de Medeiros Varzielas}
\email{ivo.de@udo.edu}
\affiliation{CFTP, Departamento de F\'{\i}sica, Instituto Superior T\'{e}cnico,
Universidade de Lisboa, Avenida Rovisco Pais 1, 1049 Lisboa, Portugal}

\author{S. F. King}
\email{king@soton.ac.uk}
\affiliation{School of Physics and Astronomy, University of Southampton,\\
SO17 1BJ Southampton, United Kingdom}

\author{Vishnudath K. N.}
\email{vishnudath.neelakand@usm.cl}
\affiliation{Departamento de F\'isica, Universidad T\'ecnica Federico Santa Mar\'ia,\\
Casilla 110-V, Valpara\'iso, Chile}

\date{\today }

\begin{abstract}
We present a unified model of quarks and leptons with modular $S_3$ flavour symmetry, where the two lightest family masses are naturally suppressed via a Pati-Salam version of the type I seesaw mechanism, mediated through heavier vector-like fermions. Majorana neutrino masses are further suppressed through a double seesaw mechanism. The viable parameter space has a preferred range of the modulus field with Im$(\tau) \sim 2 $, leading to successful fermion masses and mixing.
 The prediction for neutrinoless double beta decay is partly within the reach of the nEXO experiment. In particular, the Dirac CP violating neutrino oscillation phase is predicted to lie in the range
 $\delta_{\rm CP}^{\nu}\sim 260^o-360^o$.
\end{abstract}

\maketitle



\section{Introduction}
\label{intro}
Despite its remarkable agreement with experimental data, the current theory of strong and electroweak interactions - the standard model (SM) of particle physics - lacks an underlying mechanism to explain the strong hierarchy in the masses of elementary charged fermions. Additionally, the different mixing patterns in the quark and lepton sectors remain unexplained within the SM. The theory also fails to account for several issues, such as the tiny masses of active neutrinos and the origin of parity violation in the electroweak interaction, whose basic V-A nature is introduced by hand in the formulation of the Standard Model. This has motivated the development of several new physics models that aim to explain some or all of these unresolved issues.

Recently, the use of modular symmetries in extensions of the SM as a way of explaining the observed pattern of SM fermion masses and mixing angles has received a lot of attention from the theoretical particle physics community. See, for instance, \cite%
{Feruglio:2017spp,Ding:2023htn,Novichkov:2018nkm,King:2019vhv,Okada:2019xqk,Liu:2019khw,Kobayashi:2019rzp,Du:2020ylx,Abbas:2020qzc,Novichkov:2021evw,Ishiguro:2022pde,Chen:2023mwt,Meloni:2023aru,Kobayashi:2023qzt,Ding:2024fsf,Belfkir:2024uvj,Marciano:2024nwm}. Models based on discrete flavor symmetry, along with modular symmetry do
not include flavon fields in the particle spectrum excepting the modulus $%
\tau$, thus making the scalar sector of these theories more minimal than 
that of models not having modular symmetries. When the complex modulus $\tau$
acquires a non-vanishing vacuum expectation value ($vev$), the flavor symmetry
is spontaneously broken. Theories with modular
flavor symmetries do not require the implementation of a mechanism
responsible for the vacuum alignment, they need instead a mechanism to
determine the modulus $\tau$, which however we shall not address here. 
However, in modular flavor models, the Yukawa
couplings depend on the modular forms, which are holomorphic functions of $\tau$ \cite{Feruglio:2017spp}, which may thus be determined phenomenologically. For example, such models have been proposed with Pati-Salam unification together with $A_4$ modular symmetry \cite{Ding:2024fsf}. The lightness of the first two families is not fully addressed in many such approaches, and to remedy this the weighton mechanism has been proposed~\cite{King:2020qaj}, together with other strategies~\cite{Ding:2023htn}. Here we shall follow a different path, motivated by the type I seesaw mechanism~\cite{Minkowski:1977sc,Yanagida:1979as,GellMann:1980vs,Glashow:1979nm,Mohapatra:1979ia,Schechter:1980gr}, in order to explain the smallness of the first and second family masses. 

In this paper, in order to address the SM fermion flavor puzzle and to provide dynamical origin of the
parity violation of the electroweak interactions, we propose a 
minimal modular model based on the smallest quark-lepton unified symmetry, the Pati-Salam $SU\left( 4\right) _{C}\times
SU\left( 2\right) _{L}\times SU\left( 2\right) _{R}$ gauge group~\cite{Pati:1974yy},
and the smallest modular symmetry, $S_3$. The modular $S_3$ is perfect for implementing a two-family seesaw mechanism, since it admits doublet representations. The masses of the third family of SM
charged fermion arise from renormalisable 
Yukawa interactions involving a
colourless scalar bi-doublet as well as a bi-doublet scalar in the adjoint
representation of $SU\left( 4\right) _{C}$. The Dirac masses of the first and second families (including neutrinos) arise
from a generalised version of the type I seesaw mechanism, but applied to both charged and neutral Dirac masses~\cite{Froggatt:1978nt,Berezhiani:1983hm,Rajpoot:1987fca,Davidson:1987mh,Davidson:1987mi}. In our proposed model, the tiny active Majorana neutrino masses then arise from a double seesaw mechanism.
The model is shown to describe all quark and lepton (including neutrino) masses and mixing angles, in terms of high energy mass scales, together with complex dimensionless Yukawa coefficients which are all of order unity, and a single complex modulus field $\tau$ with Im$(\tau) \sim 2$.

In Section \ref{sec:model} we present the details of the model, followed by a numerical study of the input parameters leading to viable observables in Section \ref{sec:numerical}.
Section \ref{sec:conclusions} sumarises our findings.
A review of modular flavour symmetry can be found in Appendix \ref{sec:modular-symmetry}.

\section{The model \label{sec:model}}

We propose an extended Pati-Salam theory where the $SU\left( 4\right) _{C}\times
SU\left( 2\right) _{L}\times SU\left( 2\right) _{R}$ gauge symmetry is
supplemented by an $S_{3}$ modular symmetry. 
The masses of the third-generation SM charged fermions arise from Yukawa interactions involving the scalar bi-doublets $\Phi$ and $\Sigma$, which transform as singlet and adjoint representations of $SU(4)_C$, respectively.

The field content is enlarged by the inclusion of heavy vector-like fermions and right-handed
Majorana neutrinos, required for the implementation of the tree level
two family seesaw mechanism that yields the masses of the first and second
generation of SM charged fermions as well as the Double Seesaw mechanism that
produces the tiny masses of the light active neutrinos.
Specifically, we have vector-like fermions $\Psi _{n}$ and $\Psi _{n}^{c}$ ($n=1,2$) transforming as $%
\left( \mathbf{4,1,2}\right) $ and $\left( \overline{\mathbf{4}}\mathbf{,1,%
\overline{\mathbf{2}}}\right) $, respectively, under the Pati-Salam group.
The vector-like fermions $\Psi _{n}$ and $\Psi _{n}^{c}$ ($n=1,2$)
are the seesaw messengers which mix with the SM fermionic multiplet fields $%
F_{i}$ and $F_{i}^{c}$ ($i=1,2,3$), also transforming as $\left( \mathbf{4,1,2%
}\right) $ and $\left( \overline{\mathbf{4}}\mathbf{,1,\overline{\mathbf{2}}}%
\right) $, respectively, under the Pati-Salam group. Such mixings between SM
fermions and the seesaw messengers occurs thanks to the Yukawa interactions
involving the singlet scalar fields $\sigma _{n}$ ($n=1,2$) as well as the $%
SU\left( 4\right) _{C}$ adjoint scalars $\Xi _{1}$ and $\Xi _{2}$. Besides
that, we include three Majorana neutrinos, 
$S_{i}$ ($i=1,2,3$), which are singlets under the 
$SU(4)_{C}\otimes SU\left( 2\right) _{L}\otimes SU\left( 2\right) _{R}$
group, in order to implement the double seesaw mechanism for the generation
of light active neutrino masses.
\begin{table}[]
\begin{eqnarray*}
\begin{array}{|c|c|c|c|c|c|}\hline
& SU\left( 4\right) _{C} & SU\left( 2\right) _{L} & SU\left( 2\right) _{R} & 
S_{3} & k \\ \hline
F=\left( F_{1},F_{2}\right) & \mathbf{4} & \mathbf{2} & \mathbf{1} & \mathbf{%
2} & 0 \\ \hline
F_{3} & \mathbf{4} & \mathbf{2} & \mathbf{1} & \mathbf{1}^{\prime } & -1 \\ \hline
F_{1}^{c} & \overline{\mathbf{4}} & \mathbf{1} & \overline{\mathbf{2}} & 
\mathbf{1} & 1 \\ \hline
F_{2}^{c} & \overline{\mathbf{4}} & \mathbf{1} & \overline{\mathbf{2}} & 
\mathbf{1}^{\prime } & -1 \\ \hline
F_{3}^{c} & \overline{\mathbf{4}} & \mathbf{1} & \overline{\mathbf{2}} & 
\mathbf{1}^{\prime } & 1 \\ \hline
\Psi _{1} & \mathbf{4} & \mathbf{1} & \mathbf{2} & \mathbf{1}^{\prime } & -2
\\ \hline
\Psi _{2} & \mathbf{4} & \mathbf{1} & \mathbf{2} & \mathbf{1}^{\prime } & -4
\\ \hline
\Psi _{1}^{c} & \overline{\mathbf{4}} & \mathbf{1} & \overline{\mathbf{2}} & 
\mathbf{1}^{\prime } & 2 \\ \hline
\Psi _{2}^{c} & \overline{\mathbf{4}} & \mathbf{1} & \overline{\mathbf{2}} & 
\mathbf{1}^{\prime } & 4 \\ \hline
S^{c}=\left(S^{c}_1,S^{c}_2\right) & \mathbf{1} & \mathbf{1} & \mathbf{1} & \mathbf{2} & 1 \\ \hline
S_{3}^{c} & \mathbf{1} & \mathbf{1} & \mathbf{1} & \mathbf{1}^{\prime } & 1
\\ \hline
\Phi & \mathbf{1} & \mathbf{2} & \mathbf{2} & \mathbf{1} & 0 \\ \hline
\chi _{R} & \overline{\mathbf{4}} & \mathbf{1} & \mathbf{2} & \mathbf{1} & 2
\\ \hline
\Sigma & \mathbf{15} & \mathbf{2} & \mathbf{2} & \mathbf{1} & 0 \\ \hline
\sigma _{1} & \mathbf{1} & \mathbf{1} & \mathbf{1} & \mathbf{1} & 7 \\ \hline
\sigma _{2} & \mathbf{1} & \mathbf{1} & \mathbf{1} & \mathbf{1} & 5 \\ \hline
\Xi _{1} & \mathbf{15} & \mathbf{1} & \mathbf{1} & \mathbf{1} & 7 \\ \hline
\Xi _{2} & \mathbf{15} & \mathbf{1} & \mathbf{1} & \mathbf{1} & 5 \\ \hline
Y_{\mathbf{1}}^{\left( 4\right) }\left( \tau \right) & \mathbf{1} & \mathbf{1%
} & \mathbf{1} & \mathbf{1} & 4 \\ \hline
Y_{\mathbf{1}}^{\left( 6\right) }\left( \tau \right) & \mathbf{1} & \mathbf{1%
} & \mathbf{1} & \mathbf{1} & 6 \\ \hline
Y_{\mathbf{2}}^{\left( 2\right) }\left( \tau \right) & \mathbf{1} & \mathbf{1%
} & \mathbf{1} & \mathbf{2} & 2 \\ \hline
Y_{\mathbf{2}}^{\left( 4\right) }\left( \tau \right) & \mathbf{1} & \mathbf{1%
} & \mathbf{1} & \mathbf{2} & 4 \\ \hline
Y_{\mathbf{2}}^{\left( 6\right) }\left( \tau \right) & \mathbf{1} & \mathbf{1%
} & \mathbf{1} & \mathbf{2} & 6 \\  \hline
\end{array}
\label{Particlecontent}
\end{eqnarray*}
\vspace{-0.4cm}
    \caption{The transformation properties of the scalar and fermionic fields, as well as those of the Yukawa couplings, under the Pati-Salam gauge group and $S_3$ modular symmetry, where the modular weights of the fields are labelled by $k$.}
    \label{tab:my_label}
\end{table}
The full symmetry $\mathcal{G}$ of our model features the following
spontaneous breaking pattern: 
\begin{eqnarray}
&&\mathcal{G}=SU(4)_{C}\otimes SU\left( 2\right) _{L}\otimes SU\left(
2\right) _{R}\otimes S_{3}  \notag \\
&&\hspace{35mm}\Downarrow \Lambda _{PS}  \notag \\[0.12in]
&&SU(3)_{C}\otimes SU\left( 2\right) _{L}\otimes SU\left( 2\right)
_{R}\otimes U\left( 1\right) _{B-L}  \notag \\
&&\hspace{35mm}\Downarrow v_{R}  \notag \\[0.12in]
&&\hspace{15mm}SU(3)_{C}\otimes SU\left( 2\right) _{L}\otimes U\left(
1\right) _{Y}  \notag \\[0.12in]
&&\hspace{35mm}\Downarrow v  \notag \\[0.12in]
&&\hspace{15mm}SU(3)_{C}\otimes U\left( 1\right) _{Q}
\end{eqnarray}
where $v=246$ GeV and it is assumed that the Pati-Salam gauge symmetry is broken
at the scale $\Lambda _{PS}%
\mathrel{\rlap{\lower4pt\hbox{\hskip1pt$\sim$}}\raise1pt\hbox{$>$}}10^{6}$ GeV, which arises from the experimental bound on the branching ratio for the rare meson decays $K_L^0 \rightarrow\mu^{\pm} e^{\mp}$ mediated by the vector leptoquarks, as indicated in Refs. 
\cite{Valencia:1994cj,Smirnov:2007hv}. The $SU\left( 4\right) _{C}\times SU\left( 2\right) _{L}\times SU\left( 2\right) _{R}$ Pati-Salam symmetry is spontaneously broken down to the $SU(3)_{C}\otimes SU\left( 2\right) _{L}\otimes SU\left( 2\right)
_{R}\otimes U\left( 1\right) _{B-L}$ gauge group by the $vev$s of the scalar multiplets $\Xi_1$ and $\Xi_2$ which transform as the adjoint representation of the Pati-Salam gauge group. The second stage of symmetry breaking is triggered by the $vev$ of the scalar multiplet $\chi_R$ that transforms as a $\left( \overline{\mathbf{4}}\mathbf{,1,\mathbf{2}}\right) $ under the Pati-Salam group.
The scalars $\chi _{R}$, $\Phi $ and $\Sigma $ develop $vev$s of the form 
\begin{equation}
\left\langle \chi _{R}\right\rangle =\frac{1}{\sqrt{2}}\left( 
\begin{array}{cccc}
0 & 0 & 0 & v_{R} \\ 
0 & 0 & 0 & 0%
\end{array}%
\right) ,\hspace{1.5cm}\left\langle \Phi \right\rangle =\frac{1}{\sqrt{2}}%
\left( 
\begin{array}{cc}
v_{1} & 0 \\ 
0 & v_{2}%
\end{array}%
\right) ,\hspace{1.5cm}\left\langle \Sigma \right\rangle =\frac{1}{\sqrt{2}}%
\left( 
\begin{array}{cc}
v_{\Sigma _{1}}T^{15} & 0_{4\times 4} \\ 
0_{4\times 4} & v_{\Sigma _{2}}T^{15}%
\end{array}%
\right) ,
\end{equation}
with $T^{15}=\frac{1}{2\sqrt{6}}\mathrm{diag}\left( 1,1,1,-3\right) $.

The SM fermions can be written in component form as follows: 
\begin{equation}
F_{i}=\left( 
\begin{array}{cccc}
u_{i} & u_{i} & u_{i} & \nu _{i} \\ 
d_{i} & d_{i} & d_{i} & l_{i}%
\end{array}%
\right) ^{T},\hspace{1.5cm}F_{i}^{c}=\left( 
\begin{array}{cccc}
u_{i}^{c} & u_{i}^{c} & u_{i}^{c} & \nu _{i}^{c} \\ 
d_{i}^{c} & d_{i}^{c} & d_{i}^{c} & l_{i}^{c}%
\end{array}%
\right),\hspace{1.5cm}i=1,2,3.
\end{equation}
Similarly, the heavy vector-like fermionic multiplets $\Psi _{n}$ and $\Psi _{n}^{c}$ ($n=1,2$) containing the two family seesaw messengers are expressed as follows:
\begin{equation}
\Psi _{n}=\left( 
\begin{array}{cccc}
U_{n} & U_{n} & U_{n} & N_{n} \\ 
D_{n} & D_{n} & D_{n} & E_{n}%
\end{array}%
\right) ^{T},\hspace{1.5cm}\Psi _{n}^{c}=\left( 
\begin{array}{cccc}
U_{n}^{c} & U_{n}^{c} & U_{n}^{c} & N_{n}^{c} \\ 
D_{n}^{c} & D_{n}^{c} & D_{n}^{c} & E_{n}^{c}%
\end{array}%
\right),\hspace{1.5cm}n=1,2.
\end{equation}

The transformation properties of the scalar and fermionic fields, as well as those of the Yukawa couplings, under the Pati-Salam gauge group and $S_3$ modular symmetry are given in Table-\ref{tab:my_label}.
With this particle content and symmetries, the Yukawa superpotential compatible with the $S_3$ modular symmetry is:
\begin{eqnarray}
-\mathcal{W} &=&y_{1}Y_{\mathbf{2}}^{\left( 2\right) }\left( \tau \right)
F\Phi \Psi _{1}^{c}+y_{2}Y_{\mathbf{2}}^{\left( 4\right) }\left( \tau
\right) F\Phi \Psi _{2}^{c}+y_{3}F_{3}\Phi F_{3}^{c}\notag \\
&&+z_{1}Y_{\mathbf{1}^{\prime }}^{\left( 6\right) }\Psi _{1}\sigma
_{1}F_{1}^{c}+z_{2}Y_{\mathbf{1}}^{\left( 4\right) }\left( \tau \right) \Psi
_{1}\sigma _{1}F_{2}^{c}+z_{3}Y_{\mathbf{1}}^{\left( 4\right) }\left( \tau
\right) \Psi _{1}\sigma _{2}F_{3}^{c}+z_{4}\Psi _{2}\sigma
_{2}F_{2}^{c}+z_{5}Y_{\mathbf{1}}^{\left( 4\right) }\left( \tau \right) \Psi
_{2}\sigma _{1}F_{3}^{c}\notag \\
&&+w_{1}Y_{\mathbf{1}^{\prime }}^{\left( 6\right) }\Psi _{1}\Xi
_{1}F_{1}^{c}+w_{2}Y_{\mathbf{1}}^{\left( 4\right) }\left( \tau \right) \Psi
_{1}\Xi _{1}F_{2}^{c}+w_{3}Y_{\mathbf{1}}^{\left( 4\right) }\left( \tau
\right) \Psi _{1}\Xi _{2}F_{3}^{c}+w_{4}\Psi _{2}\Xi _{2}F_{2}^{c}+w_{5}Y_{%
\mathbf{1}}^{\left( 4\right) }\left( \tau \right) \Psi _{2}\Xi _{1}F_{3}^{c}
\notag\\
&&+x_{1}Y_{\mathbf{2}}^{\left( 2\right) }\left( \tau \right) F\Sigma \Psi
_{1}^{c}+x_{2}Y_{\mathbf{2}}^{\left( 4\right) }\left( \tau \right) F\Sigma
\Psi _{2}^{c}+x_{3}F_{3}\Sigma F_{3}^{c}+m_{\Psi _{1}}\Psi _{1}\Psi
_{1}^{c}+m_{\Psi _{2}}\Psi _{2}\Psi _{2}^{c}\notag \\
&&+\gamma _{1}Y_{\mathbf{2}}^{\left( 4\right) }\left( \tau \right)
F_{1}^{c}\chi _{R}S^{c}+\gamma _{2}Y_{\mathbf{2}}^{\left( 2\right) }\left(
\tau \right) F_{2}^{c}\chi _{R}S^{c}+\gamma _{3}Y_{\mathbf{2}}^{\left(
4\right) }\left( \tau \right) F_{3}^{c}\chi _{R}S^{c}\notag \\
&&+\gamma _{4}Y_{\mathbf{1}}^{\left( 4\right) }\left( \tau \right)
F_{3}^{c}\chi _{R}S_{3}^{c}+M_{1}Y_{\mathbf{2}}^{\left( 2\right) }\left(
\tau \right) \left( S^{c}S^{c}\right) _{\mathbf{2}}+M_{2}Y_{\mathbf{2}%
}^{\left( 2\right) }\left( \tau \right) \left( S^{c}S_{3}^{c}\right) +%
\mathrm{h.c.}. \label{superpotential}
\end{eqnarray}

\begin{figure}[H]
\begin{tabular}{cc}
\includegraphics[width=0.5\textwidth]{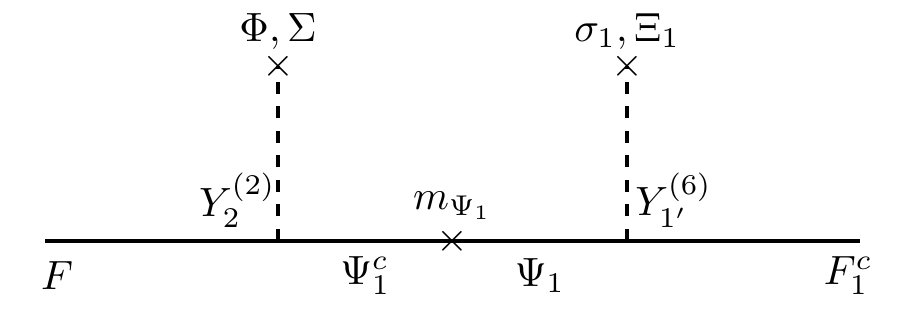} \includegraphics[width=0.5%
\textwidth]{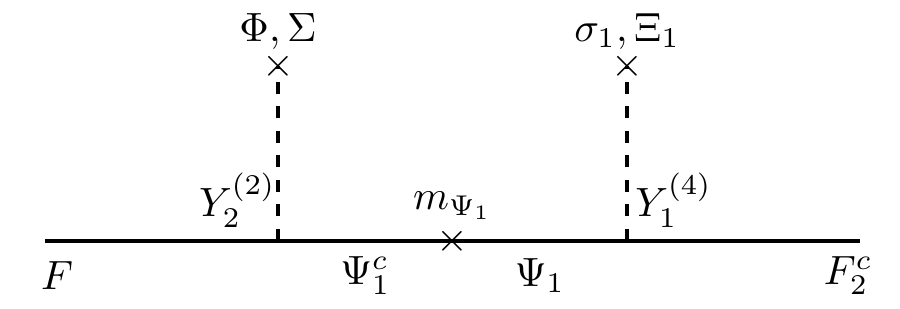} &  \\ 
\includegraphics[width=0.5\textwidth]{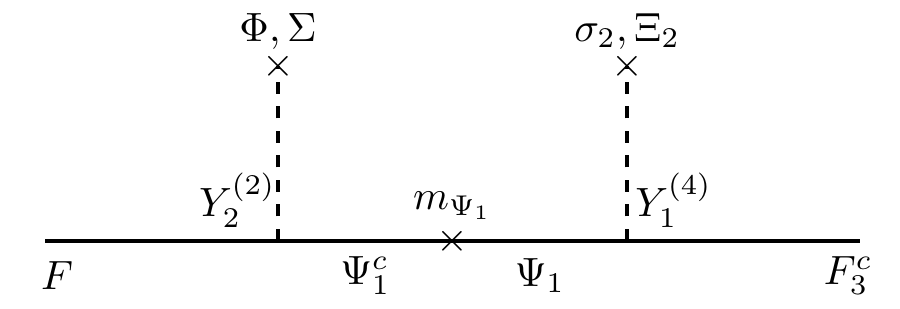} \includegraphics[width=0.5%
\textwidth]{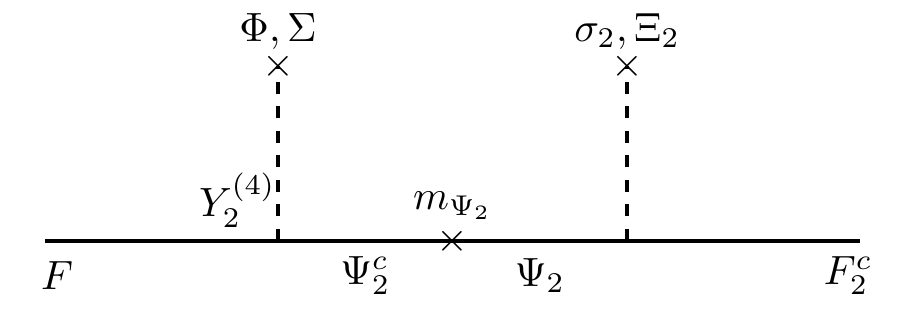} &  \\ 
\includegraphics[width=0.5\textwidth]{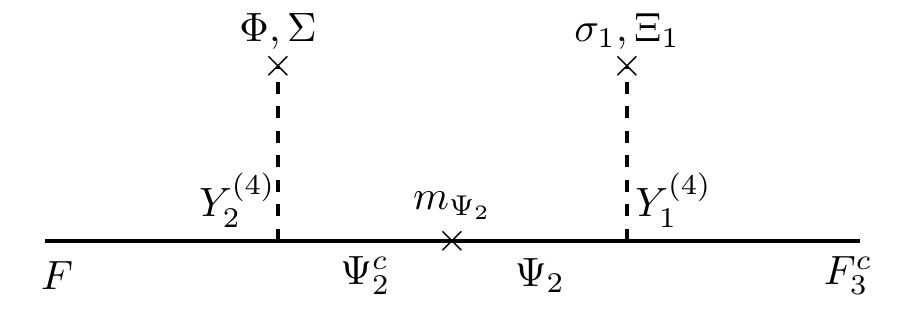} \includegraphics[width=0.325%
\textwidth]{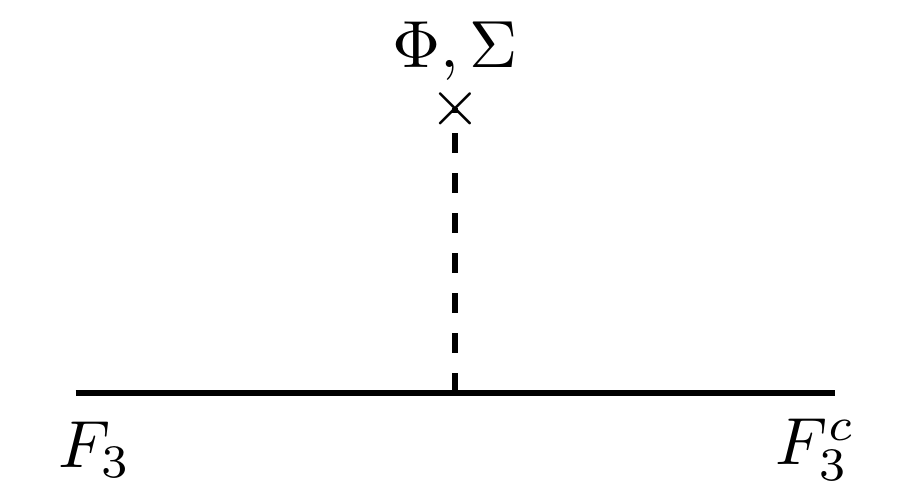} & 
\end{tabular}%
\caption{Feynman diagrams corresponding to the generation of the masses of
the charged fermions as well as for the Dirac neutrino submatrix. The charged fermions of the first and second
generations get their masses via seesaw-like diagrams mediated by $\protect\psi%
_1 $ and $\protect\psi_2$, whereas those of the third generation get their
masses through their Yukawa couplings to $\Phi$ and $\Sigma$ fields.}
\label{figCF}
\end{figure}

The flavour structure of the superpotential is replicated with $\Phi$, $\Sigma$ having the same $S_3$ assignments (being distinguished by the gauge group), and the same holds for the pairs $\sigma_1$, $\Xi_1$ and for $\sigma_2$, $\Xi_2$. The two pairs are distinct as $\sigma_1$, $\Xi_1$ have modular weight $k = 5$ whereas $\sigma_2$, $\Xi_2$ have modular weight $k=7$. This structure is also visible in the diagrams in Fig.\ref{figCF}. The structure of the diagrams is similar to the diagram of the seesaw mechanism. 
In particular, the first two families of all the charged fermions obtain their masses via seesaw mechanism mediated by the heavy vector-like fermions $\Psi_1$ and $\Psi_2$, whereas the third families obtain their masses via their Yukawa couplings to $\Phi$ and $\Sigma$. The neutrinos also obtain Dirac masses in a similar way, which is further extended to Double Seesaw by the inclusion of the singlet fields $S_i^c$.


Due to the difference in modular weights, the superpotential terms are such that the effective Yukawa terms arise with the respective modular forms, leading to 
the following mass terms for charged fermions and neutrinos:
\begin{eqnarray}
&&\left( 
\begin{array}{cc}
u_{i} & U_{k}%
\end{array}%
\right) \left( 
\begin{array}{cc}
M_{c}^{\left( u\right) } & M_{a}^{\left( u\right) } \\ 
M_{b}^{\left( u\right) } & M_{U}%
\end{array}%
\right) \left( 
\begin{array}{c}
u_{j}^{c} \\ 
U_{k}^{c}%
\end{array}%
\right) \,,\qquad \left( 
\begin{array}{cc}
d_{i} & D_{k}%
\end{array}%
\right) \left( 
\begin{array}{cc}
M_{c}^{\left( d\right) } & M_{a}^{\left( d\right) } \\ 
M_{b}^{\left( d\right) } & M_{D}%
\end{array}%
\right) \left( 
\begin{array}{c}
d_{j}^{c} \\ 
D_{k}^{c}%
\end{array}%
\right) \,, \label{quarkmatrix} \\
&&\left( 
\begin{array}{cc}
e_{i} & E_{k}%
\end{array}%
\right) \left( 
\begin{array}{cc}
M_{c}^{\left( e\right) } & M_{a}^{\left( e\right) } \\ 
M_{b}^{\left( e\right) } & M_{E}%
\end{array}%
\right) \left( 
\begin{array}{c}
e_{j}^{c} \\ 
E_{k}^{c}%
\end{array}%
\right)\,,\qquad
\left( 
\begin{array}{cc}
\nu_{i} & N_{k}%
\end{array}%
\right) \left( 
\begin{array}{cc}
M_{c}^{\left( \nu\right) } & M_{a}^{\left( \nu\right) } \\ 
M_{b}^{\left( \nu\right) } & M_{N}%
\end{array}%
\right) \left( 
\begin{array}{c}
\nu_{j}^{c} \\ 
N_{k}^{c}%
\end{array}%
\right), \label{leptonmatrix}
\end{eqnarray}


where the heavy vector-like seesaw mediator mass matrices are,
\be M_{U}=M_{D}=M_{E}=M_N = \left( 
\begin{array}{cc}
m_{\Psi _{1}} & 0 \\ 
0 & m_{\Psi _{2}}%
\end{array}%
\right). \ee

The $3 \times 3$ sub-matrices $M_c^{(u,d,e,\nu)}$ in Eqs.~\ref{quarkmatrix} and \ref{leptonmatrix} have only the $(3,3)$ element as non-zero and are given as
\begin{eqnarray}
\left( M_{c}^{\left( u\right) }\right) _{ij} &=&\frac{1}{\sqrt{2}}\left(
y_{3}v_{1}+x_{3}v_{\Sigma _{1}}\right) \delta _{i3}\delta _{j3},\hspace{0.5cm%
}\hspace{0.5cm}\left( M_{c}^{\left( d\right) }\right) _{ij}=\frac{1}{\sqrt{2}%
}\left( y_{3}v_{2}+x_{3}v_{\Sigma _{2}}\right) \delta _{i3}\delta _{j3}, \\
\left( M_{c}^{\left( e\right) }\right) _{ij} &=&\frac{1}{\sqrt{2}}\left(
y_{3}v_{2}-3x_{3}v_{\Sigma _{2}}\right) \delta _{i3}\delta _{j3},\hspace{%
0.5cm}\hspace{0.5cm}(M_{c}^{\left( \nu \right) })_{ij} =\frac{1}{\sqrt{2}}\left(
y_{3}v_{1}-3x_{3}v_{\Sigma _{1}}\right) \delta _{i3}\delta _{j3},
\label{M}
\end{eqnarray}
and these correspond to the masses of the third families of fermions generated from their Yukawa couplings to $\Phi$ and $\Sigma$. The remaining sub-matrices appearing in the quark sector are given as,
\begin{eqnarray}
M_{a}^{\left( u\right) } &=&\frac{1}{\sqrt{2}}\left( 
\begin{array}{cc}
\left( y_{1}v_{1}+x_{1}v_{\Sigma _{1}}\right) Y_{\mathbf{2,}2}^{\left(
2\right) }\left( \tau \right) & \left( y_{2}v_{1}+x_{2}v_{\Sigma
_{1}}\right) Y_{\mathbf{2,}2}^{\left( 4\right) }\left( \tau \right) \\ 
-\left( y_{1}v_{1}+x_{1}v_{\Sigma _{1}}\right) Y_{\mathbf{2,}1}^{\left(
2\right) }\left( \tau \right) & -\left( y_{2}v_{1}+x_{2}v_{\Sigma
_{1}}\right) Y_{\mathbf{2,}1}^{\left( 4\right) }\left( \tau \right) \\ 
0 & 0%
\end{array}%
\right) , \\
M_{b}^{\left( u\right) } &=& \frac{1}{\sqrt{2}}%
\left( 
\begin{array}{ccc}
Y_{\mathbf{1}^{\prime }}^{\left( 6\right) }\left( z_{1}v_{\sigma
_{1}}+w_{1}v_{\Xi _{1}}\right) & Y_{\mathbf{1}}^{\left( 4\right) }\left(
\tau \right) \left( z_{2}v_{\sigma _{1}}+w_{2}v_{\Xi _{1}}\right) & Y_{%
\mathbf{1}}^{\left( 4\right) }\left( \tau \right) \left( z_{3}v_{\sigma
_{2}}+w_{3}v_{\Xi _{2}}\right) \\ 
0 & z_{4}v_{\sigma _{2}}+w_{4}v_{\Xi _{2}} & Y_{\mathbf{1}}^{\left( 4\right)
}\left( \tau \right) \left( z_{5}v_{\sigma _{1}}+w_{5}v_{\Xi _{1}}\right)%
\end{array}%
\right) = M_{b}^{\left( d\right) } , \\
M_{a}^{\left( d\right) } &=&\frac{1}{\sqrt{2}}\left( 
\begin{array}{cc}
\left( y_{1}v_{2}+x_{1}v_{\Sigma _{2}}\right) Y_{\mathbf{2,}2}^{\left(
2\right) }\left( \tau \right) & \left( y_{2}v_{2}+x_{2}v_{\Sigma
_{2}}\right) Y_{\mathbf{2,}2}^{\left( 4\right) }\left( \tau \right) \\ 
-\left( y_{1}v_{2}+x_{1}v_{\Sigma _{2}}\right) Y_{\mathbf{2,}1}^{\left(
2\right) }\left( \tau \right) & -\left( y_{2}v_{2}+x_{2}v_{\Sigma
_{2}}\right) Y_{\mathbf{2,}1}^{\left( 4\right) }\left( \tau \right) \\ 
0 & 0%
\end{array}%
\right) ,
\end{eqnarray}
whereas those in the lepton sector are given as,
\begin{eqnarray}
M_{a}^{\left( e\right) } &=&\frac{1}{\sqrt{2}}\left( 
\begin{array}{cc}
\left( y_{1}v_{2}-3x_{1}v_{\Sigma _{2}}\right) Y_{\mathbf{2,}2}^{\left(
2\right) }\left( \tau \right) & \left( y_{2}v_{2}-3x_{2}v_{\Sigma
_{2}}\right) Y_{\mathbf{2,}2}^{\left( 4\right) }\left( \tau \right) \\ 
-\left( y_{1}v_{2}-3x_{1}v_{\Sigma _{2}}\right) Y_{\mathbf{2,}1}^{\left(
2\right) }\left( \tau \right) & -\left( y_{2}v_{2}-3x_{2}v_{\Sigma
_{2}}\right) Y_{\mathbf{2,}1}^{\left( 4\right) }\left( \tau \right) \\ 
0 & 0%
\end{array}%
\right), 
\end{eqnarray}
\begin{eqnarray}
M_{b}^{\left( e\right) } &=&\frac{1}{\sqrt{2}}\left( 
\begin{array}{ccc}
Y_{\mathbf{1}^{\prime }}^{\left( 6\right) }\left( z_{1}v_{\sigma
_{1}}-3w_{1}v_{\Xi _{1}}\right) & Y_{\mathbf{1}}^{\left( 4\right) }\left(
\tau \right) \left( z_{2}v_{\sigma _{1}}-3w_{2}v_{\Xi _{1}}\right) & Y_{%
\mathbf{1}}^{\left( 4\right) }\left( \tau \right) \left( z_{3}v_{\sigma
_{2}}-3w_{3}v_{\Xi _{2}}\right) \\ 
0 & z_{4}v_{\sigma _{2}}-3w_{4}v_{\Xi _{2}} & Y_{\mathbf{1}}^{\left(
4\right) }\left( \tau \right) \left( z_{5}v_{\sigma _{1}}-3w_{5}v_{\Xi
_{1}}\right)%
\end{array}%
\right) = M_{b}^{\left( \nu \right) },\\
M_{a}^{\left( \nu \right) } &=&\frac{1}{\sqrt{2}}\left( 
\begin{array}{cc}
\left( y_{1}v_{1}-3x_{1}v_{\Sigma _{1}}\right) Y_{\mathbf{2,}2}^{\left(
2\right) }\left( \tau \right) & \left( y_{2}v_{1}-3x_{2}v_{\Sigma
_{1}}\right) Y_{\mathbf{2,}2}^{\left( 4\right) }\left( \tau \right) \\ 
-\left( y_{1}v_{1}-3x_{1}v_{\Sigma _{1}}\right) Y_{\mathbf{2,}1}^{\left(
2\right) }\left( \tau \right) & -\left( y_{2}v_{1}-3x_{2}v_{\Sigma
_{1}}\right) Y_{\mathbf{2,}1}^{\left( 4\right) }\left( \tau \right) \\ 
0 & 0%
\end{array}%
\right).
\end{eqnarray}

Thus, once we integrate out the heavy vector like fermions $\Psi_1$ and $\Psi_2$, the masses for the first and second generation of the SM charged fermions are obtained via Seesaw mechanism, which also yields the Dirac neutrino sub-matrix, $\widetilde{M_{\nu }}$. The resulting effective low energy $3\times3$ mass matrices for the SM charged fermions as well as the Dirac neutrino matrix are: 
\begin{eqnarray}
&&\widetilde{M}_{u}=M_{c}^{\left( u\right) }-M_{a}^{\left( u\right)
}M_{U}^{-1}M_{b}^{\left( u\right) }\,,\qquad \qquad \widetilde{M}%
_{d}=M_{c}^{\left( d\right) }-M_{a}^{\left( d\right)
}M_{D}^{-1}M_{b}^{\left( d\right) }\,, \\
&&\widetilde{M}_{e}=M_{c}^{\left( e\right) }-M_{a}^{\left( e\right)
}M_{E}^{-1}M_{b}^{\left( e\right) }\,,\qquad \qquad \widetilde{M_{\nu }}=M_{c}^{\left( \nu \right) }-M_{a}^{\left( \nu
\right) }M_{N}^{-1}M_{b}^{\left( \nu \right) }\, .
\label{M2}
\end{eqnarray}

As mentioned before, our model also contains extra singlet fermions $S_i^c$ that couple to the right handed neutrinos $\nu^c$ (last two lines in Eq.\ref{superpotential}). Thus the resultant neutral fermion mass terms (after integrating out the $\Psi$ fields) can be written as,
\be
\frac{1}{2}\left( 
\begin{array}{ccc}
\nu & \nu ^{c} & S^{c}%
\end{array}%
\right)\left( 
\begin{array}{ccc}
0_{3\times 3} & \widetilde{M}_{\nu } & 0_{3\times 3} \\ 
\widetilde{M}_{\nu }^{T} & 0_{3\times 3} & M_{R} \\ 
0_{3\times 3} & M_{R}^{T} & M_{S}%
\end{array}%
\right)\left( 
\begin{array}{c}
\nu \\ 
\nu ^{c} \\ 
S^{c}%
\end{array}%
\right) + \textrm{h.c.}.  \label{Lnu}
\ee
In the above equation, all the sub-matrices are $3 \times 3$ with the Dirac mass matrix $\widetilde{M}_{\nu }$ determined by Eq.~\ref{M2}, while $M_R$ and $M_S$ are given as
\be
M_{R}=\left( 
\begin{array}{ccc}
\gamma _{1}Y_{\mathbf{2,}1}^{\left( 4\right) }\left( \tau \right) & \gamma
_{1}Y_{\mathbf{2,}2}^{\left( 4\right) }\left( \tau \right) & 0 \\ 
-\gamma _{2}Y_{\mathbf{2,}2}^{\left( 2\right) }\left( \tau \right) & \gamma
_{2}Y_{\mathbf{2,}1}^{\left( 2\right) }\left( \tau \right) & 0 \\ 
-\gamma _{3}Y_{\mathbf{2,}2}^{\left( 4\right) }\left( \tau \right) & \gamma
_{3}Y_{\mathbf{2,}1}^{\left( 4\right) }\left( \tau \right) & \gamma _{4}Y_{%
\mathbf{1}}^{\left( 4\right) }\left( \tau \right)%
\end{array}%
\right) \frac{v_{R}}{\sqrt{2}}, \,\,\,\,\,\,\,
M_{S} = \left( 
\begin{array}{ccc}
-M_{1}Y_{\mathbf{2,}1}^{\left( 2\right) }\left( \tau \right) & M_{1}Y_{%
\mathbf{2,}2}^{\left( 2\right) }\left( \tau \right) & -M_{2}Y_{\mathbf{2,}%
2}^{\left( 2\right) }\left( \tau \right) \\ 
M_{1}Y_{\mathbf{2,}2}^{\left( 2\right) }\left( \tau \right) & M_{1}Y_{%
\mathbf{2,}1}^{\left( 2\right) }\left( \tau \right) & M_{2}Y_{\mathbf{2,}%
1}^{\left( 2\right) }\left( \tau \right) \\ 
-M_{2}Y_{\mathbf{2,}2}^{\left( 2\right) }\left( \tau \right) & M_{2}Y_{%
\mathbf{2,}1}^{\left( 2\right) }\left( \tau \right) & 0%
\end{array}%
\right).
\ee


In the limit $M_S\gg M_R \gg \widetilde{M}_{\nu }$, the mass matrix in Eq.\ref{Lnu} corresponds to the double seesaw\cite{Mohapatra:1986aw},
\footnote{For a seesaw review see e.g.~\cite{King:2025eqv}.} according to which, once the heavy fields $\nu^c$ and $S^c$ are integrated out, the mass matrix for the light active neutrinos reads:
\begin{equation}
{M}_{\nu }= \bigskip \widetilde{M}_{\nu }M_{R}^{-1}M_{S}M_{R}^{-1}\widetilde{M}_{\nu }^{T}.
\end{equation}

\section{Numerical analysis \label{sec:numerical}}

In this section, we present the results of the numerical analysis conducted to evaluate the viability of the model in explaining the observed fermion masses and mixing. We vary all input parameters, including Yukawa couplings, $vev$s and masses, and minimize the function $\chi^2$, which is defined as
\be \chi^2 = \sum_{i} \left( \frac{O_{i_\text{calc}} - O_{i_\text{exp}}}{O_{i_\text{exp}}} \right )^2, \label{chisq} \ee
to determine the best fit parameters that reproduce the observed fermion masses and mixing. 

\begin{table}[H] 
  \setlength\tabcolsep{0.25cm}
  \centering
  \begin{tabular}{|c|c|}
    \hline
    Input &     \\
    Parameters & Best fit value for NH of light neutrinos   \\
    \hline
       
    $\tau$    &  $-0.03224 ~+~i~ 1.86682$   \\
    
    $m_{\Psi_1}, m_{\Psi_2}$ (GeV) & $2.38 \times 10^{12}$, $8.71 \times 10^{12}$  \\
    
    $v_1, v_2$ (GeV)      &  245.62150, 3.95665      \\ 
    
    $v_{\Sigma_1}, v_{\Sigma_2}$ (GeV) & 8.37732, 10.01214     \\
    
    $v_{\sigma_1}, v_{\sigma_2}$ (GeV) & $6.12 \times 10^{12}$, $4.64 \times 10^{12}$     \\
    
    $v_{\Xi_1}, v_{\Xi_2}$ (GeV) & $7.90 \times 10^{12}$, $3.53 \times 10^{12}$     \\ 
    
    $v_R$ (GeV)    & $ 1.23 \times 10^{11} $   \\ 
    
    $M_1, M_2$ (GeV)  & $1.0 \times 10^{12}$, $6.20 \times 10^{8}$      \\ 
    
    $x_1,x_2,x_3$  & $3.53382 ~ e^{i~ 2.23402}$, $2.93598 ~ e^{i~3.85910 }$,  $0.20033 ~ e^{i~ 2.15046} $   \\

    $y_1,y_2,y_3$  & $ 0.27120~ e^{i~ 5.93154}$, $ 0.46397~ e^{i~5.47007}$, $ 0.98750~ e^{i~ 2.24996}$  \\

    $w_1,w_2,w_3,w_4,w_5$  & $ 0.75168~ e^{i~0.96442}$, $ 0.20136~ e^{i~2.04578}$, $ 1.37663~ e^{i~5.91647}$, $ 0.23368~ e^{i~ 0.86247}$, $ 2.62639~ e^{i~5.87924 }$   \\

    $z_1,z_2,z_3,z_4,z_5$  & $ 2.37549~ e^{i~0.67600}$, $ 0.25164~ e^{i~ 1.45687}$, $ 2.46850~ e^{i~4.02943}$, $ 0.32353~ e^{i~0.15040 }$, $ 3.54392~ e^{i~6.12983}$    \\

     $\gamma_1,\gamma_2,\gamma_3,\gamma_4$  &  $ 0.20302~ e^{i~ 3.09154}$, $ 1.99746~ e^{i~ 6.05522}$, $ 3.42165~ e^{i~0.22612 }$, $ 1.96501~ e^{i~5.93289}$  
     \\
  \hline 
  $M_{\textrm{heavy neutrinos}}$ (GeV) & $ 6.06521 \times 10^5 $, $2.49460 \times 10^9$, $2.64148 \times 10^{9}$, $3.95307 \times 10^{9}$, \\ 
                                 & $1.25075 \times 10^{11}$, $1.28998 \times 10^{11}$ \\
    \hline
 & Low energy mass matrices, masses and mixing parameters  \\
    \hline
  $\widetilde{M}_u$ (GeV) &  
    $\begin{pmatrix}
-0.00018 e^{i(-2.57621)} & -0.01312 e^{i2.50106} & 0.00174 e^{i1.43459} \\
0.00428 e^{i0.66665} & -0.19406 e^{i2.37489} & -0.06547 e^{i(-2.04282)} \\
0.00000 e^{i0.00000} & 0.00000 e^{i0.00000} & -108.38236 e^{i2.24928}
\end{pmatrix}$ \\
$\widetilde{M}_d$ (GeV) & 
    $\begin{pmatrix}
0.00016 e^{i(-0.36728)} & 0.00339 e^{i0.98476} & -0.00459 e^{i(-2.87477)} \\
-0.00409 e^{i2.87558} & -0.00659 e^{i1.86300} & 0.15841 e^{i0.40887} \\
0.00000 e^{i0.00000} & 0.00000 e^{i0.00000} & -2.51228 e^{i2.21622}
\end{pmatrix}$
    \\ 
    $m_u,m_c,m_t$ (GeV) & 0.00054, 0.2670, 172.69001 \\
    $m_d,m_s,m_b$ (GeV) & 0.00120, 0.0240, 4.180  \\
    $s_{12}^q, s_{23}^q, s_{13}^q, \delta_{CP}^q$ & 
       0.2250,  0.04182, 0.00370, $65.6750\degree$  \\
       \hline
    $\widetilde{M}_e$ (GeV) &  
    $\begin{pmatrix}
0.00011 e^{i0.67714} & -0.00335 e^{i1.79184} & 0.02874 e^{i(-0.74991)} \\
-0.00261 e^{i(-2.36319)} & -0.10577 e^{i(-2.70883)} & -0.64387 e^{i2.36816} \\
0.00000 e^{i0.00000} & 0.00000 e^{i0.00000} & 0.59512 e^{i(-1.17142)}
\end{pmatrix}$
    \\ 
    ${M}_\nu$ (eV) & 
    $ \begin{pmatrix}
0.00013 e^{i0.26925} & 0.00065 e^{i(-1.25936)} & 0.00314 e^{i0.96134} \\
0.00065 e^{i(-1.25936)} & -0.02407 e^{i2.14357} & 0.01431 e^{i(-0.64736)} \\
0.00314 e^{i0.96134} & 0.01431 e^{i(-0.64736)} & 0.00000 e^{i0.00000}
\end{pmatrix}$ 
\\
$m_e, m_\mu, m_\tau$ (GeV)  & 0.00048, 0.10155, 1.77686  \\
$m_{\nu_1} (\textrm{eV}),~\Delta m_{sol}^2, ~\Delta m_{atm}^2 (\textrm{eV}{}^2)$  &  0.00276, $7.49 \times 10^{-5}$, 0.00251 \\
$s_{12}^\nu, s_{23}^\nu, s_{13}^\nu$ & 0.55497, 0.68557, 0.14883 \\

$\delta_{CP}^\nu, \alpha^M,\beta^M$  & $323.53560\degree$, $194.46228\degree$, $297.46063\degree$ \\

\hline

  \end{tabular}
  \caption{Sample best fit input parameters for NH of active neutrinos along with the corresponding values of the calculated fermion masses and mixing parameters. The $\chi^2$ value for the given point is $6.30442 \times 10^{-15}$. The low-energy mass matrices for the up and down type quarks, charged leptons, and active light neutrinos as well as the masses of the heavy neutrinos in addition to the ones coming from $\Psi_{1,2}$ are also given. For the PMNS matrix, the Majorana phase matrix is defined as $P = \textrm{diagonal} ~(1, e^{i \alpha^M/2}, e^{i \beta^M/2})$.}
  \label{bestfit}
\end{table}

In Eq.\ref{chisq}, $O_{i_\text{calc}}$ represents the model prediction, while $O_{i_\text{exp}}$ denotes the experimental best fit value. The summation is performed over the masses of charged fermions~\cite{Xing:2020ijf, ParticleDataGroup:2022pth}, the CKM mixing angles, the CKM CP phase, the PMNS mixing angles, and the mass-squared differences of light neutrinos~\cite{deSalas:2020pgw, Esteban:2024eli}. While fitting the charged fermion masses, the masses of the first two generations are fitted at $10^{12}$GeV~\cite{Xing:2020ijf}, as they arise from the seesaw mechanism, while those of the third generation are taken at the electroweak scale~\cite{ParticleDataGroup:2022pth}.
In addition, the constraint from cosmological observations on the sum of the active light neutrino masses, $\Sigma m_{i} \leq 0.12$ eV~\cite{Planck:2018vyg}, as well as the $3\sigma$ bound on the CP phase of the PMNS matrix~\cite{deSalas:2020pgw,Esteban:2024eli}, are imposed as extra conditions. In our fit, the absolute values of the Yukawa couplings are taken to be within the range $ [0.2, \sqrt{4 \pi}]$, whereas their phases are varied in the range $[0,2\pi]$. An important result is that the model only fits the Normal Ordering (NO) of the active light neutrino masses, whereas the Inverted Ordering (IO) is disfavored.

In Table \ref{bestfit}, we present a set of sample best fit parameters that reproduce the correct fermion masses and mixing along with the corresponding values of the calculated fermion masses and mixing parameters. The low-energy mass matrices for the up- and down-type quarks, charged leptons, and active light neutrinos are also given in this table. One can see that the rows of the charged fermion mass matrices satisfy a natural hierarchy as a consequence of the modular symmetry with two family seesaw, which in turn explains the observed fermion mass hierarchy.
         
\begin{figure}[H]
\begin{center}
\includegraphics[width=0.45\textwidth]{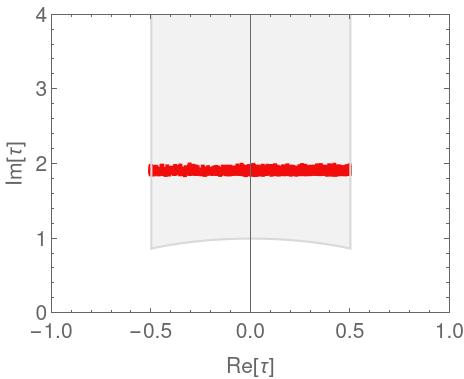} ~~~~~~
\includegraphics[width=0.45\textwidth]{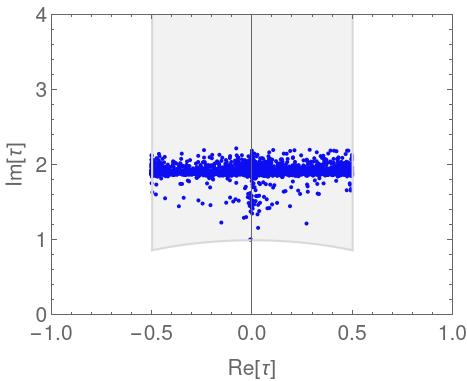}
\end{center}
\caption{The values of the real and imaginary components of the modular field $\tau$ that give $\chi^2 <= 5 \times 10^{-4}$ in our numerical scan. The fundamental domain for $\tau $ $ \left( -\frac{1}{2} \leq \textrm{Re}[\tau] \leq \frac{1}{2}, \quad |\tau| \leq 1 \right)
$ is shown by the gray-shaded region. The left panel is for fixed $vev$s and mass scales whereas they are varied in the right panel. See text for details.}
\label{figtau}
\end{figure}

In Fig.\ref{figtau}, we show the values of the real and imaginary components of the modular field $\tau$ that give $\chi^2 <= 5 \times 10^{-4}$ in our numerical scan. In the left panel, the masses of the seesaw mediators and the $vev$s are fixed according to the values provided in Table \ref{bestfit}, while the absolute values of the Yukawa couplings are varied within the range $ [0.2, \sqrt{4 \pi}]$, with their phases taking any value within $[0, 2\pi]$. In the right panel, both the mediator masses and the $vev$s are also varied freely in addition to the Yukawa couplings. The fundamental domain of
$\tau $, $ \left( -\frac{1}{2} \leq \textrm{Re}[\tau] \leq \frac{1}{2}, \quad |\tau| \geq 1 \right)$ is shown by the gray-shaded region. From the figure, one can see that the imaginary part of $\tau$ ($\tau_I$) is more restricted than the real part ($\tau_R$). This is because the magnitudes of the entries of the fermion mass matrices are more sensitive to $\tau_I$ than to $\tau_R$. This can be seen for instance, by taking the $(1,1)$ entry of $\widetilde{M}_u$ to the leading order in the $q$ - expansion of the modular forms,
\be (\widetilde{M}_u)_{11} \approx  -\frac{9 e^{-6 \pi \tau_I + 2 i \pi \tau_R} \left( e^{4 \pi \tau_I} - 16 e^{2 \pi (\tau_I + i \tau_R)} + 576 e^{4 i \pi \tau_R} \right) \left( v_{\Sigma 1} x_1 + v_1 y_1 \right) \left( v_{\Xi 1} w_1 + v_{\Sigma 1} z_1 \right)}{128 M_1},
\ee
from which we can see that $\tau_R$ always contribute to the phases  of the individual terms in the expansion. The same goes for all the mass terms. This is the reason why $\tau_I$ is restricted to be in the range $\sim [1.8,2]$ for the case of fixed mass scales and $vev$s (left panel of Fig.\ref{figtau}) as the fermion masses and mixing are more sensitive to $\tau_I$. The variation in the mass scales and the $vev$s can relax this bound because the variation in the absolute values of the mass terms due to $\tau_I$  can be compensated by taking different $vev$s and/or mass scales.

\begin{figure}[H]
\begin{center}
\includegraphics[width=0.48\textwidth]{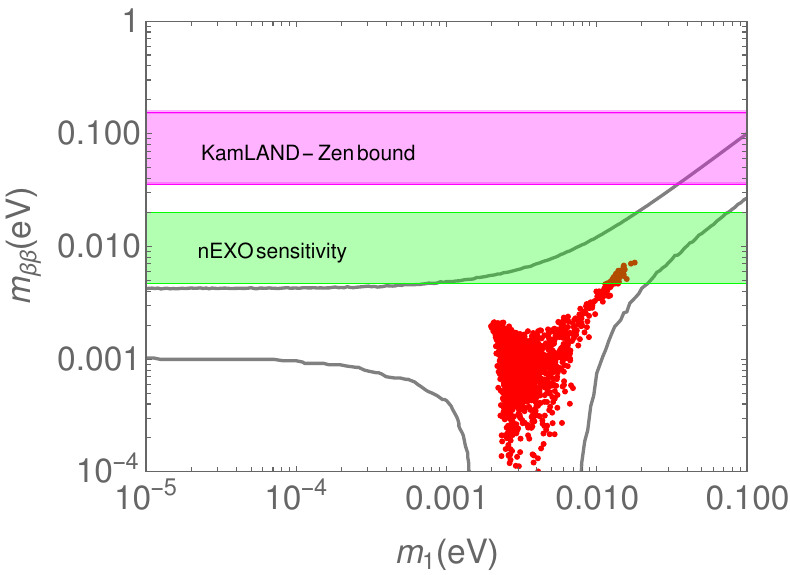} ~~~~~
\includegraphics[width=0.48\textwidth]{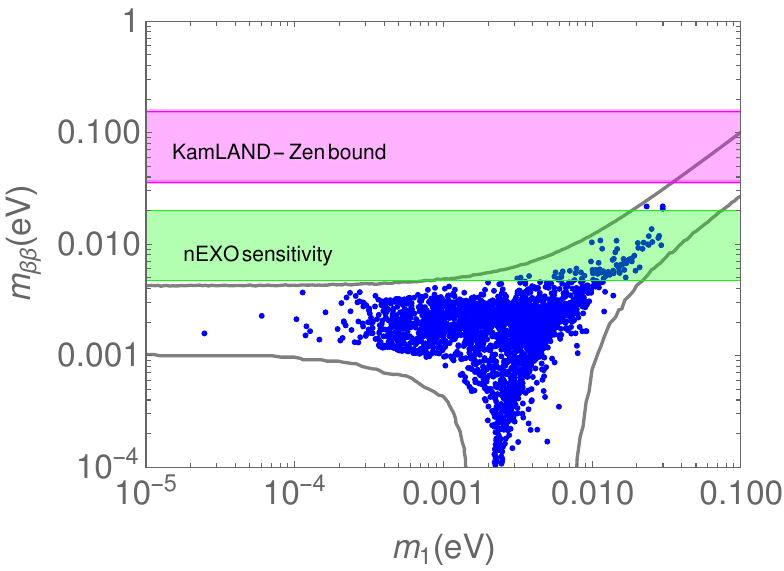} 
\end{center}
\caption{Predictions for the effective Majorana mass ($m_{ee}$) governing $0\nu\beta\beta$. The region within the solid gray lines represents the standard predictions for $m_{ee}$ assuming the active light neutrinos follow the normal hierarchy. The left and right panels correspond to the cases with fixed and varying mass scales, respectively, as in Fig. \ref{figtau}. The region above the purple band is excluded by the KamlAND-Zen experimental bound, while the green band corresponds to the projected sensitivity of the nEXO experiment.}
\label{fig0nbb}
\end{figure}

Fig.\ref{fig0nbb} shows the predictions for the effective Majorana mass ($m_{ee}$) that governs $0\nu\beta\beta$. Note that we have shown only the contributions due to the three active light neutrinos since the mixing of the heavy neutrinos with the active light neutrinos is strongly suppressed, making their contribution to $0\nu\beta\beta$ negligible.
The region within the solid gray lines represents the standard predictions for $m_{ee}$, with the assumption that the active light neutrinos follow the normal hierarchy and no modular symmetry. The red/blue points indicate the predictions from our model. As in Fig.\ref{figtau}, the masses of the seesaw mediators and the $vev$s are fixed according to the values provided in Table \ref{bestfit} in the left panel, while the magnitude and phase of the Yukawa couplings are varied within the ranges $ [0.2, \sqrt{4 \pi}]$ and $ [0, 2 \pi]$, respectively. In the right panel, both the mediator masses and the $vev$s are allowed to vary freely in addition to the Yukawa couplings. The region above the purple band is excluded by the experimental bound from KamlAND-Zen~\cite{KamLAND-Zen:2022tow}, while the green band corresponds to the projected sensitivity of the nEXO experiment~\cite{nEXO:2021ujk}. The widths of these bands are due to the uncertainty in the values of the nuclear matrix elements. One interesting feature that we can see from this figure is that the model predicts a lower bound on the mass of the lightest active neutrino. This is around $0.0025$ eV for the case of fixed mass scales while it becomes $\sim 10^{-4}$ eV in the general case. A small part of the predicted parameter space lies within the nEXO reach.

\begin{figure}[H]
\begin{center}
\includegraphics[width=0.4\textwidth]{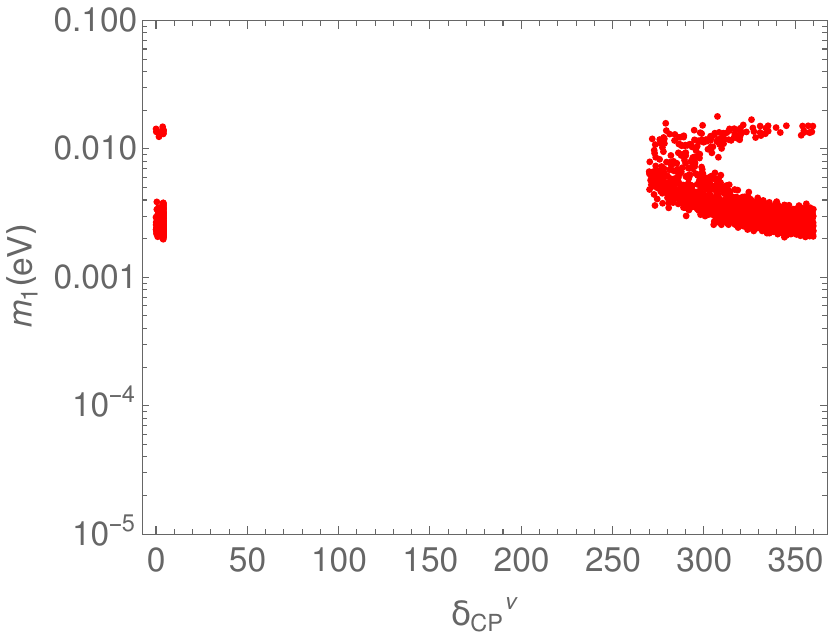}~~~~~
\includegraphics[width=0.4\textwidth]{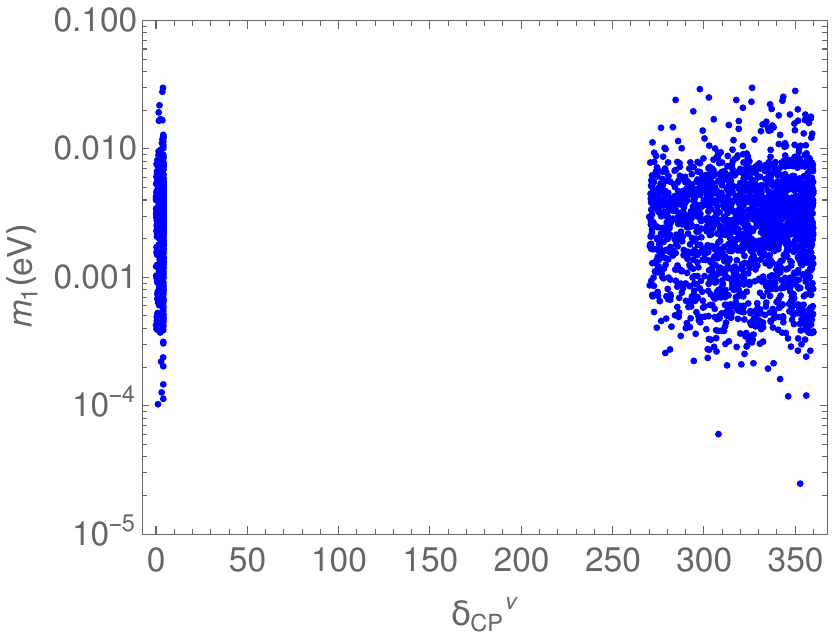} 
\includegraphics[width=0.4\textwidth]{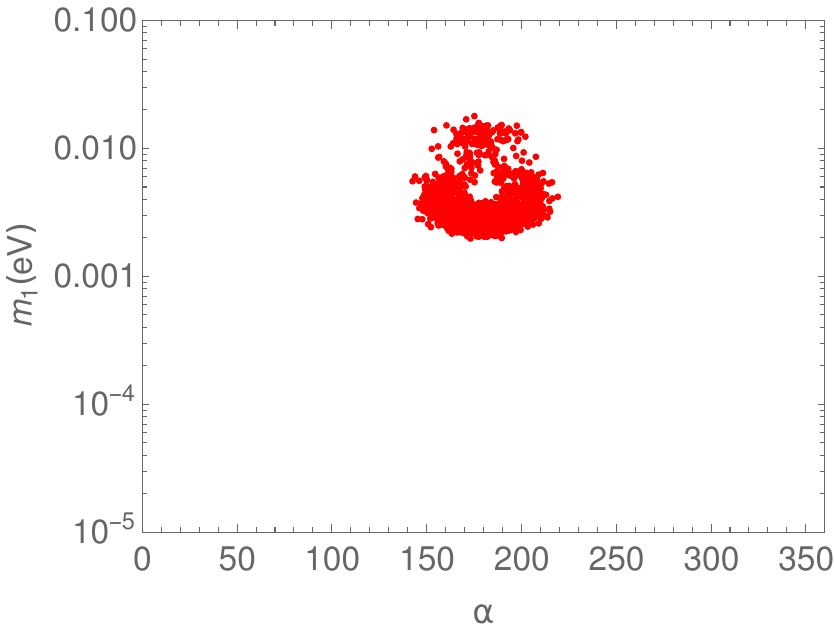}~~~~~
\includegraphics[width=0.4\textwidth]{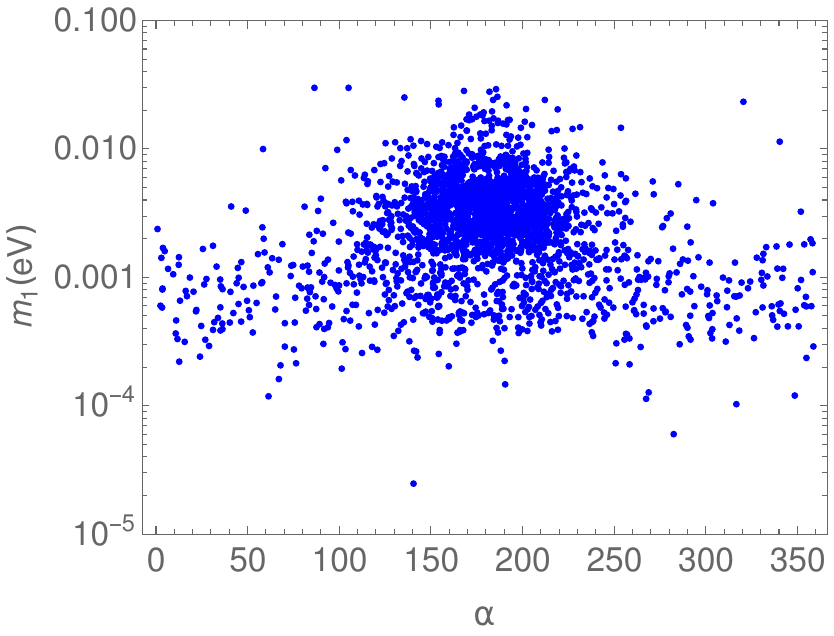} 
\end{center}
\caption{Correlations of the Dirac CP phase $\delta_{CP}^\nu$ and the Majorana phase $\alpha$ to the lightest neutrino mass $m_1$. The left and right panels correspond to the cases with fixed and varying mass scales, respectively, as in the previous figures. }
\label{phases}
\end{figure}

Fig.\ref{phases} shows the correlations of the Dirac CP phase $\delta_{CP}^\nu$ and one of the Majorana phases, $\alpha$, to the lightest neutrino mass $m_1$. As before, the left and right panels correspond to the cases with fixed and varying mass scales, respectively. It is interesting to see that for the case of fixed scales, there  exists a strong correlation between $m_1$ and $\delta_{CP}^\nu$ as well as $\alpha$, in particular for the case of fixed mass scales. Moreover, the Dirac CP violating neutrino oscillation phase is found to lie in the range $\delta_{\rm CP}^{\nu}\sim 260^o-360^o$.

\section{Conclusions \label{sec:conclusions}}

We have proposed a model based on the smallest quark-lepton unified symmetry, the Pati-Salam gauge group
and the smallest modular symmetry, $S_3$.
The masses of the third family of SM
charged fermion arise from renormalisable 
Yukawa interactions involving a
colourless scalar bi-doublet as well as a bi-doublet scalar in the adjoint
representation of $SU\left( 4\right) _{C}$. 
The first and second family masses are naturally suppressed due to a Pati-Salam version of the type I seesaw mechanism for neutrinos, but here mediated through heavier vector-like Pati-Salam fermions. Due to the Pati-Salam symmetry, the same mechanism also suppresses first and second family Dirac neutrino masses, but in the neutrino sector there are additional fields leading to tiny active Majorana neutrino masses via a Double Seesaw mechanism.

The diagrams responsible for the effective Yukawa operators are similar to those of the type I seesaw mechanism for neutrinos, with two insertions of vacuum expectation values, where one of them breaks electroweak symmetry, and one does not. The three Pati-Salam families are essentially distinguished by whether they couple to heavier vector-like fermions (the first two families) or not (third family), and this is controlled by their $S_3$ assignments.
The modular $S_3$ is perfect for implementing such a two-family seesaw mechanism, since it admits doublet representations for the first two families, and a singlet representation for the third family.

The model is shown to describe all quark and lepton (including neutrino) masses and mixing angles, in terms of high energy mass scales, together with complex dimensionless Yukawa coefficients which are all of order unity, and a single complex modulus field $\tau$ with Im$(\tau) \sim 2$. It provides a good fit to neutrino data, assuming a normal ordering of neutrino masses, while the inverted ordering is disfavoured. The associated prediction for neutrinoless double beta decay is partly within the reach of the nEXO experiment. In particular, the Dirac CP violating neutrino oscillation phase is predicted to lie in the range $\delta_{\rm CP}^{\nu}\sim 260^o-360^o$.

\section*{Acknowledgments}
A.E.C.H is supported by ANID-Chile FONDECYT 1210378, ANID-Chile FONDECYT 1241855, ANID PIA/APOYO
AFB230003 and ANID- Programa Milenio - code ICN2019\_044. V.K.N. is supported by ANID-Chile Fondecyt Postdoctoral grant 3220005. IdMV acknowledges
funding from Funda\c{c}\~{a}o para a Ci\^{e}ncia e a Tecnologia (FCT)
through the projects CFTP-FCT Unit UIDB/FIS/00777/2020 and UIDP/FIS/00777/2020, CERN/FIS-PAR/0019/2021, CERN/FIS-PAR/0002/2021, 2024.02004 CERN,
which are partially funded through POCTI (FEDER), COMPETE, QREN and EU.
A.E.C.H.
thanks the Instituto Superior T\'{e}cnico, Universidade de Lisboa for
hospitality, where part of this work was done.
S.F.K. acknowledges the STFC Consolidated Grant ST/X000583/1 and the European Union's Horizon 2020 Research and Innovation programme under the Marie Sklodowska-Curie grant agreement HIDDeN European ITN project (H2020-MSCA-ITN-2019//860881-HIDDeN).

\appendix

\section{\label{sec:modular-symmetry}Modular flavor symmetry}

In this appendix,
we provide a concise review of the main features of
modular flavor symmetry. The full modular group $\Gamma \cong \text{SL}(2,%
\mathbb{Z})$ corresponds to the group of two-dimensional matrices with
integral entries and unit determinant, 
\begin{equation}
\Gamma =\left\{ 
\begin{pmatrix}
a~ & ~b \\ 
c~ & ~d%
\end{pmatrix}%
\Big|a,b,c,d\in \mathbb{Z},~~ad-bc=1\right\} \,.
\end{equation}%
The modular group is an infinite group that can be generated by two
elements, conventionally denoted as: 
\begin{equation}
S=%
\begin{pmatrix}
0~ & ~1 \\ 
-1~ & ~0%
\end{pmatrix}%
,~~~T=%
\begin{pmatrix}
1~ & ~1 \\ 
0~ & ~1%
\end{pmatrix}%
\,,
\end{equation}%
which fulfill the relations: 
\begin{equation}
S^{4}=(ST)^{3}=1,~~S^{2}T=TS^{2}\,.  \label{eq:modu-gen}
\end{equation}%
The modular symmetry is ubiquitous in string compactifications and
corresponds to the geometrical symmetry of the extra compact space. In
simple toroidal compactification, the two-dimensional torus $T^{2}$ is
described as the quotient $T^{2}=\mathbb{C}/\Lambda _{\omega _{1},\omega
_{2}}$, where $\mathbb{C}$ stands for the whole complex plane $\mathbb{C}$
and $\Lambda _{\omega _{1},\omega _{2}}=\left\{ m\omega _{1}+n\omega
_{2},m,n\in \mathbb{Z}\right\} $ denotes a two-dimensional lattice with the
basis vectors $\omega _{1}$ and $\omega _{2}$. The lattice is left invariant
under a change in lattice basis vectors only and if only 
\begin{equation}
\begin{pmatrix}
\omega _{1} \\ 
\omega _{2}%
\end{pmatrix}%
\rightarrow 
\begin{pmatrix}
\omega _{1}^{\prime } \\ 
\omega _{2}^{\prime }%
\end{pmatrix}%
=%
\begin{pmatrix}
a~ & ~b \\ 
c~ & ~d%
\end{pmatrix}%
\begin{pmatrix}
\omega _{1} \\ 
\omega _{2}%
\end{pmatrix}%
,\qquad 
\begin{pmatrix}
a~ & ~b \\ 
c~ & ~d%
\end{pmatrix}%
\in \Gamma \,.
\end{equation}%
The torus is characterized by the complex modulus $\tau =\omega _{1}/\omega
_{2}$ up to rotation and scale transformations, without loss of generality
we can limit $\tau $ to the upper half of the complex plane with $\text{Im}(\tau
)>0 $. The two tori related by modular transformations would be identical,
i.e. 
\begin{equation}
\tau \xrightarrow{\gamma}\tau ^{\prime }=\frac{\omega _{1}^{\prime }}{\omega
_{2}^{\prime }}=\frac{a\tau +b}{c\tau +d}\equiv \gamma \tau ,~~~\text{Im}%
(\tau )>0,~~~\gamma =%
\begin{pmatrix}
a~ & ~b \\ 
c~ & ~d%
\end{pmatrix}%
\,.  \label{eq:modular-trans}
\end{equation}%
Thus the action of the generators $S$ and $T$, corresponding to modular
inversion and translation respectively, take the form: 
\begin{equation}
\tau \xrightarrow{S}-\frac{1}{\tau },~~~\tau \xrightarrow{T}\tau +1\,.
\end{equation}%
Notice that $\gamma \in \Gamma $ and $-\gamma \in \Gamma $ define the same
transformation of $\tau $. Making use of the modular transformations, it is
always possible to restrict $\tau $ to the fundamental domain defined as
follows: 
\begin{equation*}
\mathcal{D}=\left\{ \tau \,\Big|\,\text{Im}(\tau )>0,|\text{Re}(\tau )|\leq 
\frac{1}{2},|\tau |\geq 1\right\} \,.
\end{equation*}%
Any value of $\tau $ in the upper-half plane can be mapped into the
fundamental domain $\mathcal{D}$ by performing an appropriate modular
transformation, but no two points inside the fundamental domain $\mathcal{D}$
are related under the modular group. Consequently, the fundamental domain $%
\mathcal{D}$ is a representative set of the physically inequivalent modulus.
Notice that the left boundary of $\mathcal{D}$ with $\text{Re}(\tau )=-1/2$
is related to the right boundary of $\text{Re}(\tau )=1/2$ by the $T$
transformation, and the $S$ transformation maps the left unit arc $\tau
=e^{i\theta }(\pi /2\leq \theta \leq 2\pi /3)$ on the boundary is related to
the right unit arc $\tau =e^{i\theta }(\pi /3\leq \theta \leq \pi /2)$ by
the $S$ transformation.

The modular symmetry provides an origin of the discrete flavor symmetry
through the quotient, 
\begin{equation}
\Gamma_N=\Gamma/\pm\Gamma(N),~~~\Gamma^{\prime }_N=\Gamma/\Gamma(N)\,,
\end{equation}
where $\Gamma_N$ and $\Gamma^{\prime }_N$ are the inhomogeneous and
homogeneous finite modular groups respectively, and $\Gamma(N)$ is the
principal normal subgroup of level $N$, 
\begin{equation}
\Gamma(N)=\left\{%
\begin{pmatrix}
a ~ & ~ b \\ 
c ~ & ~ d%
\end{pmatrix}%
\in \Gamma, ~%
\begin{pmatrix}
a ~ & ~ b \\ 
c ~ & ~ d%
\end{pmatrix}%
=%
\begin{pmatrix}
1 ~ & ~ 0 \\ 
0 ~ & ~ 1%
\end{pmatrix}%
\;(\text{mod}~N) \right\}\,,
\end{equation}
which implies $T^{N}\in\Gamma(N)$. The inhomogeneous finite modular groups $%
\Gamma_N$ for $N=2, 3, 4, 5$ are isomorphic to the permutation groups $S_3$, 
$A_4$, $S_4$ and $A_5$ respectively~\cite%
{deAdelhartToorop:2011re,Feruglio:2017spp}, and $\Gamma^{\prime }_N$ is the
double cover of $\Gamma_N$~\cite{Liu:2019khw}.

In the framework of $\mathcal{N}=1$ global supersymmetry, the modulus $\tau $
is a chiral supermultiplet and its scalar component is restricted to the
upper half of the complex plane, and the action takes the form 
\begin{equation}
\mathcal{S}=\int d^{4}xd^{2}\theta d^{2}\bar{\theta}\,\mathcal{K}(\tau ,\bar{%
\tau};\Phi _{I},\bar{\Phi}_{I})+\left[ \int d^{4}xd^{2}\theta \,\mathcal{W}%
(\tau ,\Phi _{I})+\text{h.c.}\right] \,,
\end{equation}%
where the K\"{a}hler potential $\mathcal{K}(\tau ,\bar{\tau};\Phi _{I},\bar{%
\Phi}_{I})$ is a real gauge-invariant function of the chiral superfields $%
\tau $, $\Phi _{I}$ and their conjugates, the superpotential $\mathcal{W}%
(\tau ,\Phi _{I})$ is a holomorphic gauge invariant function of the chiral
superfields $\tau $, $\Phi _{I}$. Under the action of $\gamma \in \Gamma $,
the superfield $\Phi _{I}$ have the following non-linear transformation: 
\begin{equation}
\Phi _{I}\xrightarrow{\gamma}(c\tau +d)^{-k_{I}}\rho _{I}(\gamma )\Phi
_{I}\,,
\end{equation}%
where the weight $k_{I}$ is an integer and $\rho _{I}$ is a unitary
representation of the finite modular group $\Gamma _{N}$ or $\Gamma
_{N}^{\prime }$. The K\"{a}hler potential is assumed to take the following
minimal form 
\begin{equation}
\mathcal{K}(\tau ,\bar{\tau};\Phi _{I},\bar{\Phi}_{I})=-h\log (-i\tau +i\bar{%
\tau})+\sum_{I}\frac{\bar{\Phi}_{I}\Phi _{I}}{(-i\tau +i\bar{\tau})^{k_{I}}}%
\,,
\end{equation}%
which is invariant up to K\"{a}hler transformations. It yields the kinetic
terms of for the scalar components of $\tau $ and $\Phi _{I}$ after the
modulus acquire a $vev$.

In the concerning to the superpotential $\mathcal{W}(\tau ,\Phi _{I})$, it
can be expressed as follows 
\begin{equation}
\mathcal{W}(\tau ,\Phi _{I})=\sum_{n}Y_{I_{1}\ldots I_{n}}(\tau )\Phi
_{I_{1}}\ldots \Phi _{I_{n}}\,.
\end{equation}%
Modular invariance of $\mathcal{W}$ requires that $Y_{I_{1}\ldots
I_{n}}(\tau )$ should be a modular form of weight $k_{Y}$ and level $N$
transforming in the representation $\rho _{Y}$ of $\Gamma _{N}$ (or $\Gamma
_{N}^{\prime }$), i.e., 
\begin{equation}
Y_{I_{1}\ldots I_{n}}(\tau )\xrightarrow{\gamma}Y_{I_{1}\ldots I_{n}}(\gamma
\tau )=(c\tau +d)^{k_{Y}}\rho _{Y}(\gamma )Y_{I_{1}\ldots I_{n}}(\tau )\,.
\end{equation}%
The modular weights and the representations should fullfill the following
conditions 
\begin{equation}
k_{Y}=k_{I_{1}}+\ldots +k_{I_{n}}\,,\qquad \rho _{Y}\otimes \rho
_{I_{1}}\otimes \ldots \otimes \rho _{I_{n}}\supset \bm{1}\,,
\end{equation}%
where $\bm{1}$ denotes the trivial singlet of $\Gamma _{N}$ (or $\Gamma
_{N}^{\prime }$).

In the present work, we shall concerned with the inhomogeneous finite modular group $S_3$. The group $\Gamma _{2}\cong S_{3}$ is the permutation group of order $3$ with 6 elements, which can be expressed in terms of the two $S$ and $T$  generators satisfying the following relations~\cite{Ishimori:2010au}: 
\begin{equation}
S^{2}=T^{2}=(ST)^{3}=1\,.
\end{equation}%
The six elements of $\Gamma _{2}\cong S_{3}$ can be grouped into three
conjugacy classes 
\begin{equation}
1C_{1}=\left\{ 1\right\} ,\quad 3C_{2}=\{S,T,TST\},\quad 2C_{3}=\{ST,TS\}\,,
\end{equation}%
where $nC_{k}$ stands for the conjugacy class of $k$ elements of order $n$.
The irreducible representations of the finite modular $S_{3}$ group are two
singlets $\bm{1}$ and $\bm{1^\prime}$, and one doublet $\bm{2}$. Here we
work in the basis of diagonal matrix representation for the $T$ generator.
The representation matrices for the $S$ and $T$  generators in the three $%
S_{3}$ irreducible representations take the form: 
\begin{eqnarray}
\bm{1} &:&~~\rho _{{\bm1}}(S)=1,\qquad \rho _{{\bm1}}(T)=1\,,  \notag
\label{eq:Tp_irre} \\
\bm{1^\prime} &:&~~\rho _{{\bm1^{\prime }}}(S)=-1,\qquad \rho _{{\bm%
1^{\prime }}}(T)=-1\,,  \notag \\
\bm{2} &:&~~\rho _{{\bm2}}(S)=-\frac{1}{2}\left( 
\begin{array}{cc}
1 & ~\sqrt{3} \\ 
\sqrt{3} & ~-1 \\ 
\end{array}%
\right) ,\qquad \rho _{{\bm2}}(T)=\left( 
\begin{array}{cc}
1 & ~0 \\ 
0 & ~-1 \\ 
\end{array}%
\right) \,.
\end{eqnarray}%
The tensor product rules between the $S_{3}$ irreducible representations are
given by: 
\begin{equation}
\bm{1}\otimes \bm{1^\prime}=\bm{1^{\prime}},\qquad \bm{1^a}\otimes \bm{2}=%
\bm{2},\qquad \bm{2}\otimes \bm{2}=\bm{1}\oplus \bm{1^{\prime}}\oplus \bm{2}%
\,,  \label{eq:S3_KP}
\end{equation}%
where $a,b=0,1$ and we denote $\bm{1^0}\equiv \bm{1}$ and $\bm{1^1}\equiv %
\bm{1^\prime}$. Regarding the product of the singlet $\bm{1^{\prime}}$ with
a doublet, we have 
\begin{equation}
\bm{1^\prime}\otimes \bm{2}=\bm{2}\,\,\sim \theta \left( 
\begin{array}{c}
\varphi _{2} \\ 
-\varphi _{1}%
\end{array}%
\right) \,.
\end{equation}%
Whereas the tensor product rule of two $S_{3}$ doublets takes the form: 
\begin{equation*}
\begin{array}{lll}
\bm{2}\otimes \bm{2}=\bm{1}\oplus \bm{1^\prime}\oplus \bm{2}, & \qquad
\qquad  & \text{with}\qquad \left\{ 
\begin{array}{l}
\bm{1}\;=\theta _{1}\varphi _{1}+\theta _{2}\varphi _{2}\,, \\ 
\bm{1^\prime}=\theta _{1}\varphi _{2}-\theta _{2}\varphi _{1}\,, \\ 
\bm{2}\;=\left( 
\begin{array}{c}
\theta _{2}\varphi _{2}-\theta _{1}\varphi _{1} \\ 
\theta _{1}\varphi _{2}+\theta _{2}\varphi _{1}%
\end{array}%
\right) \,.%
\end{array}%
\right.  \\ 
\end{array}%
\end{equation*}
In the finite modular $S_{3}$ group, there are two linearly independent
modular forms of the lowest non-trivial weight 2, which can be accommodated
into a $S_{3}$ doublet $\bm{2}$ of $S_{3}$ and the doublet is given by: 
\begin{equation}
Y_{\bm{2}}^{(2)}=\left( 
\begin{array}{c}
Y_{1}(\tau ) \\ 
Y_{2}(\tau )%
\end{array}%
\right) \,.
\end{equation}
where the modular forms $Y_{1}(\tau )$ and $Y_{2}(\tau )$ take the
form \cite{Kobayashi:2018vbk}: 
\begin{eqnarray}
Y_{1}(\tau ) &=&\frac{i}{4\pi }\left[ \frac{\eta ^{\prime }(\tau /2)}{\eta
(\tau /2)}+\frac{\eta ^{\prime }((\tau +1)/2)}{\eta ((\tau +1)/2)}-8\frac{%
\eta ^{\prime }(2\tau )}{\eta (2\tau )}\right] ,  \notag
\label{eq:MF_level2_weight2} \\
Y_{2}(\tau ) &=&\frac{\sqrt{3}i}{4\pi }\left[ \frac{\eta ^{\prime }(\tau /2)%
}{\eta (\tau /2)}-\frac{\eta ^{\prime }((\tau +1)/2)}{\eta ((\tau +1)/2)}%
\right] ,
\end{eqnarray}%
Furthermore, $\eta (\tau )$ is the Dedekind function which is defined as
follows: 
\begin{equation}
\eta (\tau )=q^{1/24}\prod_{n=1}^{\infty }\left( 1-q^{n}\right) ,\qquad
q\equiv e^{2\pi i\tau }\,.
\end{equation}%
Then, the modular forms $Y_{1,2}(\tau )$ can be expressed as follows \cite{Li:2023dvm}: 
\begin{eqnarray}
Y_{1}(\tau )
&=&1/8+3q+3q^{2}+12q^{3}+3q^{4}+18q^{5}+12q^{6}+24q^{7}+3q^{8}+39q^{9}+18q^{10}\cdots ,
\notag \\
Y_{2}(\tau ) &=&\sqrt{3}%
q^{1/2}(1+4q+6q^{2}+8q^{3}+13q^{4}+12q^{5}+14q^{6}+24q^{7}+18q^{8}+20q^{9}%
\cdots ).
\end{eqnarray}%
The modular multiplets of level $N=2$ up to weight 8 are given by: 
\begin{equation}
\begin{array}{lll}
Y_{\bm{1}}^{(4)}=\left( Y_{\bm{2}}^{(2)}Y_{\bm{2}}^{(2)}\right) _{\bm{1}%
}=(Y_{\bm{2},1}^{(2)})^{2}+(Y_{\bm{2},2}^{(2)})^{2}\,, & ~Y_{\bm{2}%
}^{(4)}=\left( Y_{\bm{2}}^{(2)}Y_{\bm{2}}^{(2)}\right) _{\bm{2}}=\left( 
\begin{array}{c}
(Y_{\bm{2},2}^{(2)})^{2}-(Y_{\bm{2},1}^{(2)})^{2} \\ 
2Y_{\bm{2},1}^{(2)}Y_{\bm{2},2}^{(2)}%
\end{array}%
\right) \,, &  \\ 
Y_{\bm{1}}^{(6)}=\left( Y_{\bm{2}}^{(2)}Y_{\bm{2}}^{(4)}\right) _{\bm{1}}=Y_{%
\bm{2},1}^{(2)}Y_{\bm{2},1}^{(4)}+Y_{\bm{2},2}^{(2)}Y_{\bm{2},2}^{(4)}, & Y_{%
\bm{1^{\prime}}}^{(6)}=\left( Y_{\bm{2}}^{(2)}Y_{\bm{2}}^{(4)}\right) _{%
\bm{1^{\prime}}}=Y_{\bm{2},1}^{(2)}Y_{\bm{2},2}^{(4)}-Y_{\bm{2},2}^{(2)}Y_{%
\bm{2},1}^{(4)}\,, &  \\ 
Y_{\bm{2}}^{(6)}=\left( Y_{\bm{2}}^{(2)}Y_{\bm{1}}^{(4)}\right) _{\bm{2}%
}=\left( 
\begin{array}{c}
Y_{\bm{2},1}^{(2)}Y_{\bm{1}}^{(4)} \\ 
Y_{\bm{2},2}^{(2)}Y_{\bm{1}}^{(4)}%
\end{array}%
\right) \,, & ~Y_{\bm{1}}^{(8)}=\left( Y_{\bm{1}}^{(4)}Y_{\bm{1}%
}^{(4)}\right) _{\bm{1}}=(Y_{\bm{1}}^{(4)})^{2}\,, &  \\ 
Y_{\bm{2a}}^{(8)}=\left( Y_{\bm{1}}^{(4)}Y_{\bm{2}}^{(4)}\right) _{\bm{2}%
}=\left( 
\begin{array}{c}
Y_{\bm{1}}^{(4)}Y_{\bm{2},1}^{(4)} \\ 
Y_{\bm{1}}^{(4)}Y_{\bm{2},2}^{(4)}%
\end{array}%
\right) \,, & ~Y_{\bm{2b}}^{(8)}=\left( Y_{\bm{2}}^{(4)}Y_{\bm{2}%
}^{(4)}\right) _{\bm{2}}=\left( 
\begin{array}{c}
(Y_{\bm{2},2}^{(4)})^{2}-(Y_{\bm{2},1}^{(4)})^{2} \\ 
2Y_{\bm{2},1}^{(4)}Y_{\bm{2},2}^{(4)}%
\end{array}%
\right) \,. & 
\end{array}
\label{eq:Yw2to8}
\end{equation}

\bibliographystyle{utphys}
\bibliography{ref.bib}

\providecommand{\href}[2]{#2}\begingroup\raggedright\begin{thebibliography}{10}

\bibitem{Feruglio:2017spp}
F.~Feruglio, {\em {Are neutrino masses modular forms?}}, \href{http://dx.doi.org/10.1142/9789813238053_0012}{pp.~227--266}.
\newblock 2019.
\newblock \href{http://arxiv.org/abs/1706.08749}{{\ttfamily arXiv:1706.08749 [hep-ph]}}.

\bibitem{Ding:2023htn}
G.-J. Ding and S.~F. King, ``{Neutrino mass and mixing with modular symmetry},'' \href{http://dx.doi.org/10.1088/1361-6633/ad52a3}{{\em Rept. Prog. Phys.} {\bfseries 87} no.~8, (2024) 084201}, \href{http://arxiv.org/abs/2311.09282}{{\ttfamily arXiv:2311.09282 [hep-ph]}}.

\bibitem{Novichkov:2018nkm}
P.~P. Novichkov, J.~T. Penedo, S.~T. Petcov, and A.~V. Titov, ``{Modular A$_{5}$ symmetry for flavour model building},'' \href{http://dx.doi.org/10.1007/JHEP04(2019)174}{{\em JHEP} {\bfseries 04} (2019) 174}, \href{http://arxiv.org/abs/1812.02158}{{\ttfamily arXiv:1812.02158 [hep-ph]}}.

\bibitem{King:2019vhv}
S.~F. King and Y.-L. Zhou, ``{Trimaximal TM$_1$ mixing with two modular $S_4$ groups},'' \href{http://dx.doi.org/10.1103/PhysRevD.101.015001}{{\em Phys. Rev. D} {\bfseries 101} no.~1, (2020) 015001}, \href{http://arxiv.org/abs/1908.02770}{{\ttfamily arXiv:1908.02770 [hep-ph]}}.

\bibitem{Okada:2019xqk}
H.~Okada and Y.~Orikasa, ``{Modular $S_3$ symmetric radiative seesaw model},'' \href{http://dx.doi.org/10.1103/PhysRevD.100.115037}{{\em Phys. Rev. D} {\bfseries 100} no.~11, (2019) 115037}, \href{http://arxiv.org/abs/1907.04716}{{\ttfamily arXiv:1907.04716 [hep-ph]}}.

\bibitem{Liu:2019khw}
X.-G. Liu and G.-J. Ding, ``{Neutrino Masses and Mixing from Double Covering of Finite Modular Groups},'' \href{http://dx.doi.org/10.1007/JHEP08(2019)134}{{\em JHEP} {\bfseries 08} (2019) 134}, \href{http://arxiv.org/abs/1907.01488}{{\ttfamily arXiv:1907.01488 [hep-ph]}}.

\bibitem{Kobayashi:2019rzp}
T.~Kobayashi, Y.~Shimizu, K.~Takagi, M.~Tanimoto, and T.~H. Tatsuishi, ``{Modular $S_3$-invariant flavor model in SU(5) grand unified theory},'' \href{http://dx.doi.org/10.1093/ptep/ptaa055}{{\em PTEP} {\bfseries 2020} no.~5, (2020) 053B05}, \href{http://arxiv.org/abs/1906.10341}{{\ttfamily arXiv:1906.10341 [hep-ph]}}.

\bibitem{Du:2020ylx}
X.~Du and F.~Wang, ``{SUSY breaking constraints on modular flavor $S_{3}$ invariant SU(5) GUT model},'' \href{http://dx.doi.org/10.1007/JHEP02(2021)221}{{\em JHEP} {\bfseries 02} (2021) 221}, \href{http://arxiv.org/abs/2012.01397}{{\ttfamily arXiv:2012.01397 [hep-ph]}}.

\bibitem{Abbas:2020qzc}
M.~Abbas, ``{Fermion masses and mixing in modular A4 Symmetry},'' \href{http://dx.doi.org/10.1103/PhysRevD.103.056016}{{\em Phys. Rev. D} {\bfseries 103} no.~5, (2021) 056016}, \href{http://arxiv.org/abs/2002.01929}{{\ttfamily arXiv:2002.01929 [hep-ph]}}.

\bibitem{Novichkov:2021evw}
P.~P. Novichkov, J.~T. Penedo, and S.~T. Petcov, ``{Fermion mass hierarchies, large lepton mixing and residual modular symmetries},'' \href{http://dx.doi.org/10.1007/JHEP04(2021)206}{{\em JHEP} {\bfseries 04} (2021) 206}, \href{http://arxiv.org/abs/2102.07488}{{\ttfamily arXiv:2102.07488 [hep-ph]}}.

\bibitem{Ishiguro:2022pde}
K.~Ishiguro, H.~Okada, and H.~Otsuka, ``{Residual flavor symmetry breaking in the landscape of modular flavor models},'' \href{http://dx.doi.org/10.1007/JHEP09(2022)072}{{\em JHEP} {\bfseries 09} (2022) 072}, \href{http://arxiv.org/abs/2206.04313}{{\ttfamily arXiv:2206.04313 [hep-ph]}}.

\bibitem{Chen:2023mwt}
M.-C. Chen, S.~F. King, O.~Medina, and J.~W.~F. Valle, ``{Quark-lepton mass relations from modular flavor symmetry},'' \href{http://dx.doi.org/10.1007/JHEP02(2024)160}{{\em JHEP} {\bfseries 02} (2024) 160}, \href{http://arxiv.org/abs/2312.09255}{{\ttfamily arXiv:2312.09255 [hep-ph]}}.

\bibitem{Meloni:2023aru}
D.~Meloni and M.~Parriciatu, ``{A simplest modular S$_{3}$ model for leptons},'' \href{http://dx.doi.org/10.1007/JHEP09(2023)043}{{\em JHEP} {\bfseries 09} (2023) 043}, \href{http://arxiv.org/abs/2306.09028}{{\ttfamily arXiv:2306.09028 [hep-ph]}}.

\bibitem{Kobayashi:2023qzt}
T.~Kobayashi, T.~Nomura, H.~Okada, and H.~Otsuka, ``{Modular flavor models with positive modular weights: a new lepton model building},'' \href{http://dx.doi.org/10.1007/JHEP01(2024)121}{{\em JHEP} {\bfseries 01} (2024) 121}, \href{http://arxiv.org/abs/2310.10091}{{\ttfamily arXiv:2310.10091 [hep-ph]}}.

\bibitem{Ding:2024fsf}
G.-J. Ding, S.-Y. Jiang, S.~F. King, J.-N. Lu, and B.-Y. Qu, ``{Pati-Salam models with A$_{4}$ modular symmetry},'' \href{http://dx.doi.org/10.1007/JHEP08(2024)134}{{\em JHEP} {\bfseries 08} (2024) 134}, \href{http://arxiv.org/abs/2404.06520}{{\ttfamily arXiv:2404.06520 [hep-ph]}}.

\bibitem{Belfkir:2024uvj}
M.~Belfkir, M.~A. Loualidi, and S.~Nasri, ``{Fermion Masses and Mixing in Pati-Salam Unification with $S_3$ Modular Symmetry},'' \href{http://arxiv.org/abs/2501.00302}{{\ttfamily arXiv:2501.00302 [hep-ph]}}.

\bibitem{Marciano:2024nwm}
S.~Marciano, D.~Meloni, and M.~Parriciatu, ``{Minimal seesaw and leptogenesis with the smallest modular finite group},'' \href{http://dx.doi.org/10.1007/JHEP05(2024)020}{{\em JHEP} {\bfseries 05} (2024) 020}, \href{http://arxiv.org/abs/2402.18547}{{\ttfamily arXiv:2402.18547 [hep-ph]}}.

\bibitem{King:2020qaj}
S.~J.~D. King and S.~F. King, ``{Fermion mass hierarchies from modular symmetry},'' \href{http://dx.doi.org/10.1007/JHEP09(2020)043}{{\em JHEP} {\bfseries 09} (2020) 043}, \href{http://arxiv.org/abs/2002.00969}{{\ttfamily arXiv:2002.00969 [hep-ph]}}.

\bibitem{Minkowski:1977sc}
P.~Minkowski, ``{$\mu \to e\gamma$ at a Rate of One Out of $10^{9}$ Muon Decays?},'' \href{http://dx.doi.org/10.1016/0370-2693(77)90435-X}{{\em Phys. Lett. B} {\bfseries 67} (1977) 421--428}.

\bibitem{Yanagida:1979as}
T.~Yanagida, ``{Horizontal gauge symmetry and masses of neutrinos},'' {\em Conf. Proc. C} {\bfseries 7902131} (1979) 95--99.

\bibitem{GellMann:1980vs}
M.~Gell-Mann, P.~Ramond, and R.~Slansky, ``{Complex Spinors and Unified Theories},'' {\em Conf. Proc. C} {\bfseries 790927} (1979) 315--321, \href{http://arxiv.org/abs/1306.4669}{{\ttfamily arXiv:1306.4669 [hep-th]}}.

\bibitem{Glashow:1979nm}
S.~L. Glashow, ``{The Future of Elementary Particle Physics},'' \href{http://dx.doi.org/10.1007/978-1-4684-7197-7_15}{{\em NATO Sci. Ser. B} {\bfseries 61} (1980) 687}.

\bibitem{Mohapatra:1979ia}
R.~N. Mohapatra and G.~Senjanovic, ``{Neutrino Mass and Spontaneous Parity Nonconservation},'' \href{http://dx.doi.org/10.1103/PhysRevLett.44.912}{{\em Phys. Rev. Lett.} {\bfseries 44} (1980) 912}.

\bibitem{Schechter:1980gr}
J.~Schechter and J.~W.~F. Valle, ``{Neutrino Masses in SU(2) x U(1) Theories},'' \href{http://dx.doi.org/10.1103/PhysRevD.22.2227}{{\em Phys. Rev. D} {\bfseries 22} (1980) 2227}.

\bibitem{Pati:1974yy}
J.~C. Pati and A.~Salam, ``{Lepton Number as the Fourth Color},'' \href{http://dx.doi.org/10.1103/PhysRevD.10.275}{{\em Phys. Rev. D} {\bfseries 10} (1974) 275--289}. [Erratum: Phys.Rev.D 11, 703--703 (1975)].

\bibitem{Froggatt:1978nt}
C.~D. Froggatt and H.~B. Nielsen, ``{Hierarchy of Quark Masses, Cabibbo Angles and CP Violation},'' \href{http://dx.doi.org/10.1016/0550-3213(79)90316-X}{{\em Nucl. Phys. B} {\bfseries 147} (1979) 277--298}.

\bibitem{Berezhiani:1983hm}
Z.~G. Berezhiani, ``{The Weak Mixing Angles in Gauge Models with Horizontal Symmetry: A New Approach to Quark and Lepton Masses},'' \href{http://dx.doi.org/10.1016/0370-2693(83)90737-2}{{\em Phys. Lett. B} {\bfseries 129} (1983) 99--102}.

\bibitem{Rajpoot:1987fca}
S.~Rajpoot, ``{See-saw masses for quarks and leptons in an ambidextrous electroweak interaction model},'' \href{http://dx.doi.org/10.1142/S0217732387000422}{{\em Mod. Phys. Lett. A} {\bfseries 2} no.~5, (1987) 307--315}. [Erratum: Mod.Phys.Lett.A 2, 541 (1987)].

\bibitem{Davidson:1987mh}
A.~Davidson and K.~C. Wali, ``{Universal Seesaw Mechanism?},'' \href{http://dx.doi.org/10.1103/PhysRevLett.59.393}{{\em Phys. Rev. Lett.} {\bfseries 59} (1987) 393}.

\bibitem{Davidson:1987mi}
A.~Davidson and K.~C. Wali, ``{SU(5)-L x SU(5)-R HYBRID UNIFICATION},'' \href{http://dx.doi.org/10.1103/PhysRevLett.58.2623}{{\em Phys. Rev. Lett.} {\bfseries 58} (1987) 2623}.

\bibitem{Valencia:1994cj}
G.~Valencia and S.~Willenbrock, ``{Quark - lepton unification and rare meson decays},'' \href{http://dx.doi.org/10.1103/PhysRevD.50.6843}{{\em Phys. Rev. D} {\bfseries 50} (1994) 6843--6848}, \href{http://arxiv.org/abs/hep-ph/9409201}{{\ttfamily arXiv:hep-ph/9409201}}.

\bibitem{Smirnov:2007hv}
A.~D. Smirnov, ``{Mass limits for scalar and gauge leptoquarks from K(L)0 ---\ensuremath{>} e-+ mu+-, B0 ---\ensuremath{>} e-+ tau+- decays},'' \href{http://dx.doi.org/10.1142/S0217732307024401}{{\em Mod. Phys. Lett. A} {\bfseries 22} (2007) 2353--2363}, \href{http://arxiv.org/abs/0705.0308}{{\ttfamily arXiv:0705.0308 [hep-ph]}}.

\bibitem{Mohapatra:1986aw}
R.~N. Mohapatra, ``{Mechanism for Understanding Small Neutrino Mass in Superstring Theories},'' \href{http://dx.doi.org/10.1103/PhysRevLett.56.561}{{\em Phys. Rev. Lett.} {\bfseries 56} (1986) 561--563}.

\bibitem{King:2025eqv}
S.~F. King, {\em {Right-handed neutrinos: seesaw models and signatures}}.
\newblock 2, 2025.
\newblock \href{http://arxiv.org/abs/2502.07877}{{\ttfamily arXiv:2502.07877 [hep-ph]}}.

\bibitem{Xing:2020ijf}
Z.-z. Xing, ``{Flavor structures of charged fermions and massive neutrinos},'' \href{http://dx.doi.org/10.1016/j.physrep.2020.02.001}{{\em Phys. Rept.} {\bfseries 854} (2020) 1--147}, \href{http://arxiv.org/abs/1909.09610}{{\ttfamily arXiv:1909.09610 [hep-ph]}}.

\bibitem{ParticleDataGroup:2022pth}
{\bfseries Particle Data Group} Collaboration, R.~Workman {\em et~al.}, ``{Review of Particle Physics},'' \href{http://dx.doi.org/10.1093/ptep/ptac097}{{\em PTEP} {\bfseries 2022} (2022) 083C01}.

\bibitem{deSalas:2020pgw}
P.~F. de~Salas, D.~V. Forero, S.~Gariazzo, P.~Mart\'\i{}nez-Mirav\'e, O.~Mena, C.~A. Ternes, M.~T\'ortola, and J.~W.~F. Valle, ``{2020 global reassessment of the neutrino oscillation picture},'' \href{http://dx.doi.org/10.1007/JHEP02(2021)071}{{\em JHEP} {\bfseries 02} (2021) 071}, \href{http://arxiv.org/abs/2006.11237}{{\ttfamily arXiv:2006.11237 [hep-ph]}}.

\bibitem{Esteban:2024eli}
I.~Esteban, M.~C. Gonzalez-Garcia, M.~Maltoni, I.~Martinez-Soler, J.~a.~P. Pinheiro, and T.~Schwetz, ``{NuFit-6.0: Updated global analysis of three-flavor neutrino oscillations},'' \href{http://arxiv.org/abs/2410.05380}{{\ttfamily arXiv:2410.05380 [hep-ph]}}.

\bibitem{Planck:2018vyg}
{\bfseries Planck} Collaboration, N.~Aghanim {\em et~al.}, ``{Planck 2018 results. VI. Cosmological parameters},'' \href{http://dx.doi.org/10.1051/0004-6361/201833910}{{\em Astron. Astrophys.} {\bfseries 641} (2020) A6}, \href{http://arxiv.org/abs/1807.06209}{{\ttfamily arXiv:1807.06209 [astro-ph.CO]}}. [Erratum: Astron.Astrophys. 652, C4 (2021)].

\bibitem{KamLAND-Zen:2022tow}
{\bfseries KamLAND-Zen} Collaboration, S.~Abe {\em et~al.}, ``{Search for the Majorana Nature of Neutrinos in the Inverted Mass Ordering Region with KamLAND-Zen},'' \href{http://dx.doi.org/10.1103/PhysRevLett.130.051801}{{\em Phys. Rev. Lett.} {\bfseries 130} no.~5, (2023) 051801}, \href{http://arxiv.org/abs/2203.02139}{{\ttfamily arXiv:2203.02139 [hep-ex]}}.

\bibitem{nEXO:2021ujk}
{\bfseries nEXO} Collaboration, G.~Adhikari {\em et~al.}, ``{nEXO: neutrinoless double beta decay search beyond 10$^{28}$ year half-life sensitivity},'' \href{http://dx.doi.org/10.1088/1361-6471/ac3631}{{\em J. Phys. G} {\bfseries 49} no.~1, (2022) 015104}, \href{http://arxiv.org/abs/2106.16243}{{\ttfamily arXiv:2106.16243 [nucl-ex]}}.

\bibitem{deAdelhartToorop:2011re}
R.~de~Adelhart~Toorop, F.~Feruglio, and C.~Hagedorn, ``{Finite Modular Groups and Lepton Mixing},'' \href{http://dx.doi.org/10.1016/j.nuclphysb.2012.01.017}{{\em Nucl. Phys. B} {\bfseries 858} (2012) 437--467}, \href{http://arxiv.org/abs/1112.1340}{{\ttfamily arXiv:1112.1340 [hep-ph]}}.

\bibitem{Ishimori:2010au}
H.~Ishimori, T.~Kobayashi, H.~Ohki, Y.~Shimizu, H.~Okada, and M.~Tanimoto, ``{Non-Abelian Discrete Symmetries in Particle Physics},'' \href{http://dx.doi.org/10.1143/PTPS.183.1}{{\em Prog. Theor. Phys. Suppl.} {\bfseries 183} (2010) 1--163}, \href{http://arxiv.org/abs/1003.3552}{{\ttfamily arXiv:1003.3552 [hep-th]}}.

\bibitem{Kobayashi:2018vbk}
T.~Kobayashi, K.~Tanaka, and T.~H. Tatsuishi, ``{Neutrino mixing from finite modular groups},'' \href{http://dx.doi.org/10.1103/PhysRevD.98.016004}{{\em Phys. Rev. D} {\bfseries 98} no.~1, (2018) 016004}, \href{http://arxiv.org/abs/1803.10391}{{\ttfamily arXiv:1803.10391 [hep-ph]}}.

\bibitem{Li:2023dvm}
C.-C. Li and G.-J. Ding, ``{Eclectic flavor group~$\Delta(27)\rtimes S_3$ and lepton model building},'' \href{http://dx.doi.org/10.1007/JHEP03(2024)054}{{\em JHEP} {\bfseries 03} (2024) 054}, \href{http://arxiv.org/abs/2308.16901}{{\ttfamily arXiv:2308.16901 [hep-ph]}}.

\end{thebibliography}\endgroup

\end{document}